\documentclass[preprintnumbers,a4paper,aps,pra,showpacs,superscriptaddress,tightenlines,twocolumn,showkeys,amsmath]{revtex4-1}
\usepackage[left=15mm,top=15mm,right=15mm,bottom=15mm]{geometry}
\usepackage{graphicx}
\usepackage{slashed} 
\usepackage{array} 
\usepackage{calligra}
\usepackage{braket}
\usepackage{float}
\usepackage{simplewick}
\usepackage{commath}

\newcommand{\be}{\begin{equation}}
\newcommand{\ee}{\end{equation}}

\newcommand{\momfrac}{\left<x\right>}
\newcommand{\zero}{\vec{0}}
\newcommand{\p}{\vec{p}}
\newcommand{\pp}{\vec{p}^{\, \prime}}
\newcommand{\q}{\vec{q}}
\newcommand{\x}{\vec{x}}
\newcommand{\xp}{\x^{\, \prime}}
\newcommand{\y}{\vec{y}}
\newcommand{\Dm}{\Delta m}
\newcommand{\ma}{m^{\alpha}}
\newcommand{\mb}{m^{\beta}}
\newcommand{\Zap}{Z^{\alpha}_{\p}}
\newcommand{\Zapbar}{\overline{Z}\,^{\alpha}_{\p}}

\newcommand{\Zappbar}{\overline{Z}\,^{\alpha}_{\pp}}
\newcommand{\Zbp}{Z^{\beta}_{\p}}

\newcommand{\Ep}{E_{\p}}
\newcommand{\Epp}{E_{\pp}}
\newcommand{\DEp}{\Delta \Ep}

\newcommand{\Eap}{E^{\alpha}_{\p}}
\newcommand{\Ebp}{E^{\beta}_{\p}}
\newcommand{\Eapp}{E^{\alpha}_{\pp}}
\newcommand{\sumx}{\sum_{\x}}
\newcommand{\sumxp}{\sum_{\xp}}

\newcommand{\sumy}{\sum_{\y}}
\newcommand{\sumxy}{\sum_{\x ,\y}}

\newcommand{\sumalpha}{\sum_{\alpha}}
\newcommand{\sumab}{\sum_{\alpha ,\beta}}
\newcommand{\lbxrb}{\left( x \right)}
\newcommand{\lbxprb}{\left( x,\p \right)}
\newcommand{\lbxpprb}{\left( x,\pp \right)}
\newcommand{\lbxtrb}{\left( \x,t \right)}
\newcommand{\lbyrb}{\left( y \right)}
\newcommand{\lbytaurb}{\left( \y,\tau \right)}

\newcommand{\lbprb}{\left( \p \right)}

\newcommand{\lbpprb}{\left( \pp \right)}

\newcommand{\lbzerorb}{\left(0\right)}
\newcommand{\lbzeroprb}{\left(0,\p\right)}

\newcommand{\lbqsqrdrb}{( q^{2} )}

\newcommand{\Lvac}{\langle \, \Omega \, | }
\newcommand{\Rvac}{| \, \Omega \, \rangle }

\newcommand{\chixt}{\chi \lbxtrb}

\newcommand{\chizerobar}{\overline{\chi}\, \lbzerorb}

\newcommand{\JOytau}{\mathcal{J}_{O} \lbytaurb}
\newcommand{\JOq}{\mathcal{J}_{O} \lbqsqrdrb}

\newcommand{\Oytau}{O \lbytaurb}

\newcommand{\phibxp}{\phi^{\beta} \lbxprb}
\newcommand{\phibxpp}{\phi^{\beta} \lbxpprb}

\newcommand{\phibzerobarp}{\overline{\phi}\,^{\beta} \lbzeroprb}

\newcommand{\psiax}{\psi^{a}_{\alpha} \lbxrb}

\newcommand{\psibybar}{\overline{\psi}\,^{b}_{\beta}  \lbyrb}

\newcommand{\fsprop}{\sigma\left(\Gamma;\pp,t;y;0\right)}
\newcommand{\fspropdag}{\sigma^{\dagger}\left(\Gamma;\pp,t;y;0\right)}

\newcommand{\Gtwopt}{G_{2} \left( \Gamma ; \p , t \right)}
\newcommand{\Gtwoptmass}{G_{2} \left( \Gamma=\Gamma_{4} ; \p , t \right)}
\newcommand{\Gtwoptij}{\left(G_{2}\right)_{ij} \left( \Gamma ; \p , t \right)}

\newcommand{\Gtwoptbeta}{G_{2}^{\beta} \left(\Gamma ;  \p , t  \right)}

\newcommand{\Gtwoptto}{G_{2} \left( \Gamma ; \p , t_{0} \right)}
\newcommand{\Gtwopttodt}{G_{2} \left( \Gamma ; \p , t_{0}+ \Delta t \right)}
\newcommand{\Gtwoptpdt}{G_{2} \left( \Gamma ; \p , t + \Delta t \right)}
\newcommand{\Gtwoptng}{G_{2} \left( \p , t \right)}
\newcommand{\Gtwopptng}{G_{2} \left( \pp , t \right)}

\newcommand{\Gtwopptngbeta}{G^{\beta}_{2} \left( \pp , t \right)}
\newcommand{\Gtwoptaung}{G_{2} \left( \p , \tau \right)}
\newcommand{\Gtwopptaung}{G_{2} \left( \pp , \tau \right)}
\newcommand{\Gtwopttaung}{G_{2} \left( \p , t-\tau \right)}
\newcommand{\Gtwoppttaung}{G_{2} \left( \pp , t-\tau \right)}

\newcommand{\GtwoptGfour}{G_{2} \left( \Gamma_{4} ; \p , t \right)}

\newcommand{\Gtwoztng}{G_{2} \left( \zero , t  \right)}

\newcommand{\Gatwoptng}{G^{\alpha}_{2} \left( \p , t \right)}

\newcommand{\Gbtwopptng}{G^{\beta}_{2} \left( \pp , t \right)}
\newcommand{\Gbtwoptaung}{G^{\beta}_{2} \left( \p , \tau \right)}
\newcommand{\Gbtwopptaung}{G^{\beta}_{2} \left( \pp , \tau \right)}
\newcommand{\Gbtwopttaung}{G^{\beta}_{2} \left( \p , t-\tau \right)}
\newcommand{\Gbtwoppttaung}{G^{\beta}_{2} \left( \pp , t-\tau \right)}

\newcommand{\Gthreept}{G_{3} \left( \Gamma ;  \pp , t ; \q, \tau ; O \right)}
\newcommand{\Gthreeppzt}{G_{3} \left( \Gamma ;  \zero , t ; \q, \tau ; O \right)}
\newcommand{\Gthreezt}{G_{3} \left( \Gamma ;  \zero , t ; \zero ,  \tau ; O \right)}

\newcommand{\Gthreeptbeta}{G_{3}^{\beta} \left( \Gamma ;  \pp , t ; \q, \tau ; O \right)}
\newcommand{\Gthreeptij}{\left(G_{3}\right)_{ij} \left( \Gamma ;  \pp , t ; \q, \tau ; O \right)}

\newcommand{\Gthreeptjbeta}{\left(G_{3}\right)^{\beta}_{j} \left( \Gamma ;  \pp , t ; \q, \tau ; O \right)}

\newcommand{\Rfactor}{R\left(\Gamma ; \pp , t ; \q, \tau ; O \right)}

\newcommand{\Rfactorbeta}{R^{\beta}\left(\Gamma ; \pp , t ; \q, \tau ; O \right)}

\newcommand{\FF}{FF\left(\Ebp \rightarrow \Eapp , \Gamma , O\right)}
\newcommand{\FFground}{FF\left(\Ep^{0} \rightarrow \Epp^{0} , \Gamma , O\right)}

\newcommand{\Fpx}{e^{-i \p \cdot \x}}

\newcommand{\Fppx}{e^{-i \pp \cdot \x}}
\newcommand{\Fppxp}{e^{-i \pp \cdot \xp}}

\newcommand{\Fqy}{e^{i \q \cdot \y}}

\newcommand{\eEt}{e^{ -\Eap t}}
\newcommand{\eEppttau}{e^{ -\Eapp \left(t - \tau \right) }}
\newcommand{\eEptau}{e^{ -\Ebp \tau }}

\newcommand{\vbip}{v^{\beta}_{i}\lbprb}
\newcommand{\vbjp}{v^{\beta}_{j}\lbprb}
\newcommand{\ubip}{u^{\beta}_{i}\lbprb}
\newcommand{\ubjp}{u^{\beta}_{j}\lbprb}
\newcommand{\vbipp}{v^{\beta}_{i}\lbpprb}

\begin{document}

\preprint{    ADP-16-23/T978}
\preprint{    Edinburgh 2016/07}
\preprint{    DESY 16-099}
\preprint{    LTH 1087}



\title{Nucleon matrix elements using the variational method in lattice QCD}


\author{J.~Dragos}
\affiliation{CSSM, Department of Physics, The University of Adelaide, Adelaide SA 5005, Australia}
\author{R.~Horsley}
\affiliation{School of Physics and Astronomy, University of Edinburgh, Edinburgh EH9 3JZ, UK}
\author{W.~Kamleh}
\affiliation{CSSM, Department of Physics, The University of Adelaide, Adelaide SA 5005, Australia}
\author{D.~B.~Leinweber}
\affiliation{CSSM, Department of Physics, The University of Adelaide, Adelaide SA 5005, Australia}
\author{Y.~Nakamura}
\affiliation{RIKEN Advanced Institute for Computational Science, Kobe, Hyogo 650-0047, Japan}
\author{P.~E.~L.~Rakow}
\affiliation{Theoretical Physics Division, Department of Mathematical Sciences, University of Liverpool, Liverpool L69 3BX, UK }
\author{G.~Schierholz}
\affiliation{Deutsches Elektronen-Synchrotron DESY, 22603 Hamburg, Germany}
\author{R.~D.~Young}
\affiliation{CSSM, Department of Physics, The University of Adelaide, Adelaide SA 5005, Australia}
\affiliation{CoEPP, Department of Physics, University of Adelaide, Adelaide SA 5005, Australia}
\author{J.~M.~Zanotti}
\affiliation{CSSM, Department of Physics, The University of Adelaide, Adelaide SA 5005, Australia}


\thanks{CSSM/QCDSF/UKQCD}


\date{\today}

\begin{abstract}
  The extraction of hadron matrix elements in lattice QCD using the
  standard two- and three-point correlator functions
  demands careful attention to systematic uncertainties.
  One of the most commonly studied sources of systematic
  error is contamination from excited states. We
  apply the variational method to calculate the axial vector current
  \(g_{A}\), the scalar current \(g_{S}\) and the quark momentum fraction
  \(\momfrac\) of the nucleon and we compare the results to the more commonly
  used summation and two-exponential fit methods. The results
  demonstrate that the variational approach offers a more efficient
  and robust method for the determination of nucleon matrix elements.

\end{abstract}

\pacs{11.15.Ha, 12.38.Gc, 14.20.Dh }
\keywords{Lattice QCD, Matrix elements, Nucleon axial charge, Variational method, Summation method, Two state fit}

\maketitle
\setlength{\parskip}{1em}
\section{\label{Intro}Introduction}
Modern lattice QCD simulations are making significant advances towards
the direct comparison with experimental results for a range of
hadronic observables. Therefore there is an increasing demand on
numerical studies to quantify all uncertainties, both statistical and
systematic. In this present work, we focus specifically on the
systematic uncertainty associated with excited-state contamination in
baryon matrix elements.
The presence of the weak signal-to-noise behaviour makes the study
of baryon 3-point functions particularly sensitive to excited-state
contamination.
In practice, there is a persistent trade-off to keep the source--sink
time separation short enough to provide a statistically significant
signal, while desiring a long enough separation to suppress excited
states.

In this study we investigate a range of techniques for
addressing excited-state contamination in baryon matrix elements.
Our focus is on the variational method, which has seen tremendous
success in spectroscopy studies \cite{Engel:2010my,Edwards:2011jj,Mahbub:2010rm,Kiratidis:2015vpa,Mahbub:2013ala,Menadue:2011pd,Blossier:2009kd}, in addition to some applications in hadonic matrix elements \cite{Owen:2015fra,Hall:2014uca,Owen:2015gva,Owen:2012ts,Bulava:2011yz,Lin:2008qv,Lin:2010fv,Yoon:2016dij}. We then compare the variational method to the popular
``two-exponential fit'' and ``summation'' methods seen in the literature \cite{Capitani:2012gj,Green:2014xba,Capitani:2015sba,Bali:2014nma,Capitani:2015sba,Bali:2014nma,Bhattacharya:2013ehc,Dinter:2011sg,Lin:2008qv,Lin:2010fv,Yoon:2016dij}.
%
%
The observables we choose to study are: the nucleon axial vector
charge $g_A$, the nucleon scalar charge $g_{S}$ and the quark
momentum fraction $\langle x\rangle$ for the nucleon. The latter two have previously
been identified as being particularly sensitive to excited-state
contamination.
The results of our analysis demonstrates, for all three quantities
considered, that the variational method offers improved reliability in
comparison to the summation and two-exponential fit methods.
%


The structure of this paper is as follows: Section \ref{Lattice}
contains an explanation of the gauge field configurations used along
with our method for creating correlation functions; Section \ref{CMT}
outlines the application of the variational approach to 3-point functions,
including a prescription for optimising the sequential source through the sink inversion;
Section \ref{SMandTSF} summarises the implementation of the summation
method and two-exponential fit; Section \ref{Res} presents the
numerical results from this paper; Section \ref{Ana} summarises our
findings and discusses the contrasting features of the various
techniques presented; and Section \ref{Con} provides concluding
remarks and future outlook.

\section{\label{Lattice}Lattice Details}
\subsection{\label{SimDet}Simulation Details}
Simulations were performed on a \(32^3 \times 64\) dimensional ensemble with a pion mass of 460 MeV and a lattice
spacing of 0.074 fm \cite{Bornyakov:2015eaa,Bietenholz:2010jr,Bietenholz:2011qq}. This
ensemble corresponds to the SU(3)-symmetric point, where
\(m_{u}=m_{d}=m_{s}\) with $\kappa = 0.120900$; which has been tuned
to be close to the physical, average light-quark mass \(\overline{m} = \frac{1}{3}\left(m_{u}+m_{d}+m_{s}\right)\) \cite{Bietenholz:2011qq}. The simulation
uses a clover action comprising of a stout smeared fermion action
along with the tree-level Symanzik improved gluon action. We perform \(\mathcal{O}(1000)\) measurements on \(\mathcal{O}(1800)\) trajectories, with multiple source location to remove autocorrelations. The
renormalization constants \(Z_{A} = 0.8728(6)(27)\) and
\(Z_{S}^{\overline{MS}} = 0.682(6)(18)\) at 2 GeV have been reported in
Ref.~\cite{Constantinou:2014fka},
whereas \(\momfrac\) remains unrenormalised in the present work.

A fixed boundary condition in Euclidean time dimension and periodic
boundary conditions in the spatial dimensions are chosen for this
calculation. As outlined in the next section, we employ the sequential source through the sink method to compute three-point functions (see \cite{Can:2015hda}). Hence we are required to fixed the sink momentum \(\pp\) for which we set \(\pp = \zero\). The space of all Hermitian matrices combined with zero and one
derivative operators has been calculated as they require minimal
computational time after the sequential propagators have been created. Although different transfer momenta has
been calculated with the zero sink momentum, this paper only analyses forward matrix elements ( zero momentum transfer ) and the three particular operators and spin
projectors corresponding to \(g_{A}\), \(\momfrac\) and
\(g_{S}\) as described in Section \ref{Res}.

The smearings undertaken in later sections are a gauge-invariant
Gaussian smearing which has the functional form \cite{Gusken:1989qx}:

\be \label{eq:Hsmear}
\begin{aligned}
&H\left(\xp ,\x \right) = \left( 1-\alpha\right) \delta_{\x \xp} +\ldots\\
&\frac{\alpha}{6}\sum_{\hat{i}} \left\lbrace \delta_{\x\left(\xp-\hat{i}\right)} \mathcal{U}_{\hat{i}}\left(\x\right)+\delta_{\x\left(\xp+\hat{i}\right)} \mathcal{U}^{\dagger}_{\hat{i}}\left(\x-\hat{i}\right)\right\rbrace,
\end{aligned}
\ee
and is applied iteratively to the source and sink quark field.

We take \(\alpha=0.7\) and then by repeated
application of this smearing operator \(N_{smear}\) times we
generate quark source and sink distributions of different spatial
sizes. To form our variational basis we solved our quark propagators
for 32, 64 and 128 sweeps of smearing which correspond to root mean square radii of 0.248 fm, 0.351 fm and 0.496 fm respectively.

To get an extensive range of source--sink separation times for the study of the summation method,
we have performed the sequential-source inversions at source-sink separations of 10, 13,
16, 19 and 22 time slices. In physical units, this corresponding to the range 0.74-1.63 fm.
This extended range is primarily at our reference source smearing of 32.
The full ensemble of inversions performed in this study are indicated in
Table~\ref{tab:inversions}.
\begin{table}
  \begin{ruledtabular}
  \begin{tabular}{rr|ccccc}
                & $t$ & 10 & 13 & 16 & 19 & 22 \\
$N_{\rm smear}$ &  &    &     &    &    &   \\
\hline
32              &  & $\star$  & $\star$   & $\star$  & $\star$  & $\star$ \\
64              &  &    & $\star$   &    &    &   \\
128             &  &    & $\star$   &    &    &   \\
variational     &  &    & $\star$   & $\star$  &    &   \\
\end{tabular}
\end{ruledtabular}
\caption{\label{tab:inversions} Table showing all smearing and source-sink separations undertaken in this paper.}
\end{table}




\subsection{\label{CfunDef}Two-Point and Three-Point Correlation Functions}
We follow standard notation for a nucleon two-point correlation
function with momentum \(\vec{p}\) at Euclidean time $t$:
\be
\Gtwopt = \sumx \Fpx \mathrm{Tr} \left\lbrace  \Gamma \Lvac \chixt \chizerobar \Rvac \right\rbrace,
\ee
where \(\chi\) is a proton interpolating operator and \(\Gamma =
\Gamma_{4} \equiv \frac{I + \gamma_{4}}{2}\) is used to project onto
positive parity states. This equation reduces to the following:
\be\label{eq:C2red} \Gtwoptng \equiv \Gtwoptmass = \sumalpha \eEt
\Zapbar \Zap, \ee where \(\Zap\) and \(\Zapbar\) are momentum dependent
constants of state \(\alpha\) related to the coupling strengths of the
operators to their energy eigenstates of energy \(\Eap\).  For
3-point correlators, we repeat with an inserted current operator
\(\Oytau\) at some intermediate time \(\tau\):

\be
\begin{split}
&\Gthreept = \sumxy \Fppx \Fqy \\
&\quad\times \mathrm{Tr} \left\lbrace \Gamma \Lvac \chixt \Oytau \chizerobar \Rvac \right\rbrace.
\end{split}
\ee

In this notation, \(\pp\) is the momentum of the final state, \(\p\) is the momentum of the initial state and the momentum transferred to the nucleon by the operator O is defined as \(\q \equiv \pp - \p \).

Reducing the three-point correlator in a similar way to the two-point
correlator, Eq.(\ref{eq:C2red}), we have:
\be
\begin{split}
&\Gthreept = \\
&\sumab \eEppttau \eEptau \Zappbar \Zbp \FF,
\end{split}
\ee
defining the ``FF'' function as:
\be
\begin{split}
&\FF \equiv \\
&\mathrm{Tr} \left\lbrace \Gamma \left(\frac{\slashed{p}' + \ma}{2\Eapp}\right) \JOq  \left(\frac{\slashed{p} + \mb}{2\Ebp}\right)  \right\rbrace,
\end{split}
\ee
where \(\Eapp\) and \(\Ebp\) are the source and sink energies, respectively, referring to the state indices \(\alpha\) and \(\beta\), with momenta \(\pp\) and \(\p\). \(\JOq\) is the appropriate form factor combination for the particular operator \(\mathcal{O}\) chosen. For example, choosing \(\mathcal{O} = i\gamma_{3}\gamma_{5}\) corresponds to:

\be
\mathcal{J}_{i\gamma_{3}\gamma_{5}}(q^{2}) = i\gamma_{5}\gamma_{3}G_{A}(q^{2}) - \frac{\gamma_{5}q_{3}}{2m}G_{P}(q^{2}).
\ee
where \(G_{A}\) and \(G_{P}\) are the axial and induced pseudo scalar form factors, respectively.

Once \(G_{2}\) and \(G_{3}\) are obtained, we can define the combination to remove the exponential time dependence and wave function overlap factors:

\be \label{eq:Rfun}
\begin{split}
  \Rfactor \equiv  \frac{\Gthreept}{\Gtwopptng} \times\\
  \sqrt{\frac{\Gtwopptaung\Gtwopptng\Gtwopttaung}{\Gtwoptaung\Gtwoptng\Gtwoppttaung}}.
\end{split}
\ee

Due to the exponential time dependence in the two- and three-point correlators, ground state dominance will occur at large times \(\tau \gg 0\) and \(t \gg \tau\). Hence, the ``FF'' function can be extracted by taking large \(\tau\) and \(t\) limits:
\be \label{eq:Rfac}
\begin{split}
  \Rfactor &\xrightarrow{\tau\gg 0,t \gg \tau}\\
  A\left(\p,\pp,m^{\alpha},m^{\beta}\right)&\FF ,
\end{split}
\ee
where \(A\left(\p,\pp,m^{\alpha},m^{\beta}\right)\) is a known kinematical constant.

\section{\label{CMT}Variational Method}
The previous section we showed how to determine the ground state properties by studying the large time behaviour of two- and three-point correlation functions. As is well known, the signal-to-noise ratio of nucleon correlation functions decreases significantly at large times. Hence with finite statistics, it is often necessary to find a balance between large source-current-sink time separations and quality of signal. To help alleviate this problem, it would be advantageous if one were able to reduce the contributions from excited states at early times in order to facilitate the extraction of ground state properties at early times.
The variational method has proven to be a robust and useful tool for studying two-point correlators in this respect \cite{Engel:2010my,Edwards:2011jj,Mahbub:2010rm,Kiratidis:2015vpa,Mahbub:2013ala,Menadue:2011pd,Blossier:2009kd}. Recently, this approach has been extended to three-point correlators, specifically aiming to reduce the effect of excited state contamination in hadronic matrix elements \cite{Owen:2015fra,Hall:2014uca,Owen:2015gva,Owen:2012ts,Bulava:2011yz,Lin:2008qv,Lin:2010fv,Yoon:2016dij}.

Once a basis of states is obtained that contains different
couplings to different energy levels, a variational analysis can be
undertaken to produce correlation functions that couple strongly to
the ground state. Given the significant signal/noise problem for
baryon correlators, any reduction in the time required to saturate the
ground state can give significant advantage in the study of 3-point
correlators.

We present our notation for the variational approach, following a
format similar to that described in Ref.~\cite{Owen:2012ts}. Ideally, the improved two-point correlation function isolating the generic state
$\beta$ is given by
\be \label{eq:CM2pt}
\Gtwoptbeta = \sumx \Fpx \mathrm{Tr} \left\lbrace  \Gamma \Lvac \phibxp \phibzerobarp \Rvac \right\rbrace,
\ee
where \(\phibxp\) and \(\phibzerobarp\) are constructed as a linear
combination of our basis of operators:
\be
\phibxp = \sum_{i} \vbip \chi_{i}\lbxrb,
\ee
\be
\phibzerobarp = \sum_{i} \ubip \overline{\chi}_{i}\lbzerorb,
\ee

If we express the correlators \( G_{2}\) created over a basis \(i\)\(j\) as a matrix of correlators, we can rewrite Eq.(\ref{eq:CM2pt}) as:
\be \label{eq:CM2ptConst}
\Gtwoptbeta = \vbip \Gtwoptij \ubjp,
\ee
which constructs a new two-point correlator that has a stronger coupling to state \(\beta\).
By selecting two sink times \(t=t_{0}\) and \(t=t_{0} + \Delta t\). \(u\) and \(v\) can be found via the solution to the following eigenvalue equations:
\be
\vbip \big[ \Gtwopttodt \Gtwoptto^{-1}\big]_{ij} = \vbjp \lambda^{\beta},
\ee
\be
\big[ \Gtwoptto^{-1}\Gtwopttodt \big]_{ij}\ubjp = \ubip \lambda^{\beta}.
\ee

For the ground state (\(\beta = 0\)), this creates a two-point function that has an accelerated approach to the ground state over euclidean time.
For this analysis, \(\left(G_{2}\right)_{ij}\) is a 3x3 matrix corresponding to 32, 64, and 128 sweeps of smearing at the source (index \(i\)) and the sink (index \(j\)).
The same \(u\) and \(v\) found for the two-point correlators at a particular momentum can be used to estimate the 3-point correlator for state $\beta$:
\be
\begin{split}
\Gthreeptbeta = \sumxy \Fppx \Fqy \\
\mathrm{Tr} \left\lbrace \Gamma \Lvac \phibxpp \JOytau \phibzerobarp \Rvac \right\rbrace,
\end{split}
\ee
or rewritten over \(i\)\(j\) as:
\be
\begin{split}
&\Gthreeptbeta = \\
\vbipp &\Gthreeptij \ubjp.
\end{split}
\ee

And lastly, we construct the same ratio as previously described in Eq.(\ref{eq:Rfac}) which will have the ``FF'' function dependence:

\be\label{eq:Rfactor}
\begin{split}
  \Rfactorbeta \equiv  \frac{\Gthreeptbeta}{\Gtwopptngbeta}\times\\
  \sqrt{\frac{\Gbtwopptaung\Gbtwopptng\Gbtwopttaung}{\Gbtwoptaung\Gatwoptng\Gbtwoppttaung}}.
\end{split}
\ee

For the following results, a set of \(t_{0} \) and \(\Delta t \) were analysed, and \(t_{0} = 2\) and \(\Delta t = 2\) were chosen, however minimal variation was observed for other choices as seen in Figure~\ref{gAVarCompGraph} in Section \ref{Res:Axial}.


\subsection{\label{FSm}Smearing the Sink}
As the variational approach we employ uses different levels of quark smearing to form our basis of operators, we first describe how to perform the standard method for smearing the sink of a three-point function before outlining our procedure for applying the variational method at the sink.
Gaussian gauge invariant smearings are applied to the source and
sink of the two-point correlation function as well as the source of the three-point
correlation function. To produce an equivalent smearing at the sink for the
three-point correlation function, a new construction is needed as the fixed sink
method does not have direct access to the operator/interpolating field
at the sink.

Two-point quark propagators are defined as:
\be
\contraction{S^{ab}_{\alpha\beta}\left(x;y\right) = }{\psi}{\left(x\right)}{\psi^{b}_{\beta}}
S^{ab}_{\alpha\beta}\left(x;y\right) = \psiax \psibybar,
\ee
where \(\overline{\psi}\) and \(\psi\) are the quark creation and annihilation operators, respectively. Hence the construction for the fixed sink method is as follows. First we write the three-point function in terms of quark propagators

\be
\begin{split}
\Gthreept = \sumy \Fqy \\
\mathrm{Tr} \left\lbrace \fsprop \mathcal{O}\left(\y,\tau\right) S\left(y;0\right)\right\rbrace,
\end{split}
\ee
where \(\sigma\) is created by solving the linear equation:

\be \label{eq:SInv}
\sumy S^{-1}\left(x;y\right)\gamma_{5}\fspropdag = e^{i\pp\cdot\x}\gamma_{5}S^{\dagger}_{C}\left(x;0\right),
\ee
with an appropriate choice of \(S_{C}\left(x;0\right)\).
The source for the inversion, \(S_{C}\) (known as a ``sequential source''), is the combination of all the quark propagators from the source to the sink that have no current operators attached to them.

\begin{figure}
\includegraphics[clip,width=.48\textwidth]{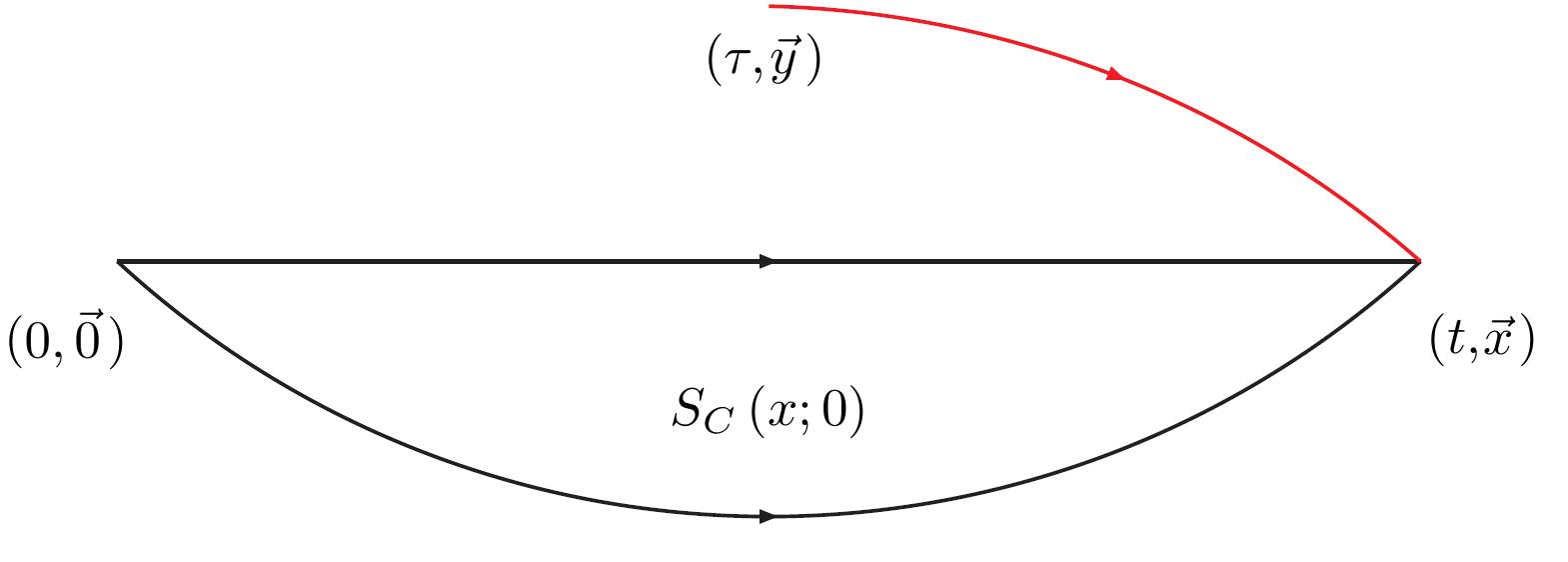}
\caption{\label{seqsource}Diagram showing \(\fsprop\) which is the sequential source propagator. \(S_{C}\left(x;0\right)\) is the source used in the inversion, which is represented by the black lines in the diagram.}
\end{figure}

To smear the sink properly, the term \( S^{-1}\left(x,y\right) \) must be smeared at the sink as well, but we can use the same inversion calculation by not applying the smearings to this term and instead smear the source \(S_{C}\) to compensate:

\be
\Fppx S_{C}\left(x;0\right) \Rightarrow \sumxp \Fppxp S_{C}\left(\xp,t;0\right)H\left(\xp,\x\right),
\ee
where \(H\) is our smearing operator used to smear the source or sink of a propagator \(S\). For this paper, a gauge invariant Gaussian smearing is undertaken as shown in Eq.(\ref{eq:Hsmear}).


\subsection{\label{VSm}Variational Method Sink Smearing}

Since in most cases a single \(\pp\) is chosen (usually \(\pp=\zero\)), we can reduce the computation time for the three-point correlator from \(n^{2}\) to \(n\) where \(n\) is the number of source and sink smearings. This is done by constructing a three-point correlator as a combination of sink smearings with weights \(v\) created from the variational method on the two-point correlators:

\be
\Gthreeptjbeta \equiv \vbipp \Gthreeptij.
\ee

So when we create the fixed sink propagator \(\sigma\), we can solve Eq.(\ref{eq:SInv}) with a smearing substitution of:

\be
\begin{aligned}
&\Fppx S_{C}\left(\pp;x;0\right) \Rightarrow \\
&\sum_{\xp,i} \Fppxp S_{C}\left(\xp,t;0\right)H_{i}\left(\xp,\x\right)\vbipp,
\end{aligned}
\ee
where \(H_{i}\) is the smearing operator applied the amount of times corresponding to basis index \(i\) (e.g. \(i=1\) might correspond to applying 32 sweeps of smearing) and \(\vbipp\) is the weightings obtained from the variational method applied to the two-point correlators.

An important point to note here is that a single combination of \(t_{0}\) and \(\Delta t\) must be chosen from the two-point correlator as \( \vbipp \) is now used in the matrix inversion calculation to create the fixed sink propagator/correlator and is dependent on these parameters.
\\
\section{\label{SMandTSF}Summation and Two-Exponential Fit Methods}
Two alternative methods that have been proposed for reducing the effect of excited state contamination in hadronic matrix element calculations are the summation and two-state fit methods.
\subsection{\label{SM}Summation Method}
As has been used many times in the past and in recent works \cite{Capitani:2012gj,Green:2014xba,Capitani:2015sba,Bali:2014nma}, a summation method can be employed in this calculation to reduce the excited state contamination. The process proceeds by summing the ratio \(\Rfactor\) over operator insertion times, \(\tau\):
\begin{widetext}
\be \label{eq:Sum}
S\left(\Gamma;\pp,t;\q;\mathcal{O}\right) = \sum_{\tau=\delta t}^{t-\delta t} \Rfactor \\
\rightarrow c+t\left\lbrace \FFground + \mathcal{O}\left(e^{-min\left(\Delta E_{\p},\Delta E_{\pp}\right) t}\right)\right\rbrace,
\ee
\end{widetext}
where \(\Delta E_{\p}\) is the energy difference between the ground
and first excited state energies with momentum \(\p\). The (apparent) advantage of this technique is
that the correction to the matrix element is suppressed by an
exponential in $t$, the full source--sink separation time. This is in
contrast to the conventional method where the parametric suppression
of excited states in given by a similar exponential of time $t-\tau$
(or $\tau$), which is $\sim t/2$ in the plateau region.
We allow for the slight generalisation of including a \(\delta t \) parameter, also considered in \cite{Bali:2014nma} which describes the number of current insertion results of the summation of $R$ which have been removed closest to both the source and sink.
This region has the strongest statistical signal, yet provides minimal information on the ground-state matrix element. In most instances, we find the results to be largely insensitive to $\delta t$, as one might expect. But the summation method results shown later for \(\momfrac \) [Figure~\ref{giDiSummary}] is an example where we see a statistically significant change when we vary the \(\delta t\) parameter.

After performing simulations at multiple source-sink separation
times, $t$, one performs a linear fit to determine $\FFground$.


\subsection{\label{TSF}Two-Exponential Fit Method}
Multi-exponential fits have also been suggested as a way of removing excited state contamination from the determination of ground-state quantities. While proposed long ago for spectroscopy, many recent studies have attempted this in hadron matrix
element calculations
\cite{Capitani:2015sba,Bali:2014nma,Bhattacharya:2013ehc,Dinter:2011sg,Lin:2008qv,Lin:2010fv,Yoon:2016dij}. For
comparative purposes, we also explore the use of a two-exponential
fit. This is undertaken by expanding the two-point and three-point functions to the second
energy state and fitting to obtain the parameters of interest. Since
all calculations performed as a part of this work have \( \pp = 0 \), the formalism can be reduced
to fitting the following functions:

\be
\Gtwoptng = \abs{Z_{\Ep}}^{2} e^{-\Ep t} + \abs{Z_{\Ep^{\prime}}}^{2} e^{-\Ep^{\prime} t} + \ldots,
\ee
\be
\begin{aligned}
&\Gthreeppzt = \\
&\overline{Z}_{\Ep}Z_{m}e^{-m t}e^{-\left(\Ep-m\right) \tau}FF\left(\Ep\rightarrow m,\Gamma,\mathcal{O}\right)\\
+&\overline{Z}_{\Ep^{\prime}}Z_{m} e^{-m t}e^{-\left(\Ep^{\prime}-m\right)\tau}FF\left(\Ep^{\prime}\rightarrow m,\Gamma,\mathcal{O}\right)\\
+&\overline{Z}_{\Ep}Z_{m^{\prime}} e^{-m^{\prime} t}e^{-\left(\Ep-m^{\prime}\right)\tau}FF\left(\Ep\rightarrow m^{\prime},\Gamma,\mathcal{O}\right)\\
+&\overline{Z}_{\Ep^{\prime}}Z_{m^{\prime}}e^{-m^{\prime} t}e^{-\left(\Ep^{\prime}-m^{\prime}\right)\tau}FF\left(\Ep^{\prime}\rightarrow m^{\prime},\Gamma,\mathcal{O}\right) \ldots,
\end{aligned}
\ee
where \(\Ep\) and \(m\) now refer to the ground state energy and mass while the primes in \(\Ep^{\prime}\) and \(m^{\prime}\) denote the first excited state energy and mass.
Taking this framework, we can fit the nucleon two-point function to the following function to determine the mass (with \(\DEp \equiv \Ep^{\prime} - \Ep\) and \(\Delta m \equiv m^{\prime} - m\)):
\be
\Gtwoptng = A_{\Ep}e^{-\Ep t} + A_{\Ep^{\prime}}e^{-\left(\Ep+\DEp\right)t},
\ee
and we can fit the nucleon three-point function by the following function from which we are then able to extract the ``FF'' function:
\be
\begin{aligned}
\Gthreeppzt = \sqrt{A_{\Ep}A_{m}}e^{-m t}e^{-\left(\Ep-m\right)\tau}\\
\left\lbrace B_{00}+B_{10}e^{-\DEp \tau}+ B_{01}e^{-\Dm \left(t-\tau\right)}\right.\\
\left. +B_{11}e^{-\Dm t}e^{-\left(\DEp-\Dm\right)\tau}\right\rbrace,
\end{aligned}
\ee
where we have 4 free parameters in the two-point correlator for each
momentum used, as well as 4 free parameters in the three-point
correlator fit which correspond to:
\begin{align}
B_{00} & = FF\left(\Ep\rightarrow m,\Gamma,\mathcal{O}\right),\\
B_{10} & = \sqrt{\frac{A_{\Ep^{\prime}}}{A_{\Ep}}}FF\left(\Ep^{\prime}\rightarrow m,\Gamma,\mathcal{O}\right),\\
B_{01} & = \sqrt{\frac{A_{m^{\prime}}}{A_{m}}}FF\left(\Ep\rightarrow m^{\prime},\Gamma,\mathcal{O}\right),\\
B_{11} & = \sqrt{\frac{A_{\Ep^{\prime}}A_{m^{\prime}}}{A_{\Ep}A_{m}}}FF\left(\Ep^{\prime}\rightarrow m^{\prime},\Gamma,\mathcal{O}\right).
\end{align}
For the forward matrix elements considered in this work we require only \(\p = \pp = 0\), which
implies \(\Ep = m\) and \(\DEp = \Dm \), and hence:
\be
\Gtwoztng = A_{m}e^{-m t} + A_{m^{\prime}}e^{-\left(m+\Delta m\right)t},
\ee
\be \label{eq:TSF}
\begin{split}
&\Gthreezt = A_{m}e^{-m t} \ldots\\
&\left\lbrace B_{0}+B_{1}\left(e^{-\Delta m \tau}+ e^{-\Delta m \left(t-\tau\right)}\right)+B_{2}e^{-\Delta m t}\right\rbrace.
\end{split}
\ee

Now there are only 3 free parameters for the three-point correlator due to the transition \(m\rightarrow m^{\prime}\) being interchangeable with \(m^{\prime}\rightarrow m\) :
\begin{align}
B_{0} &= FF\left(m\rightarrow m,\Gamma,\mathcal{O}\right),\\
B_{1} &= \sqrt{\frac{A_{m^{\prime}}}{A_{m}}}FF\left(m^{\prime}\rightarrow m,\Gamma,\mathcal{O}\right),\\
B_{2} &= \frac{A_{m^{\prime}}}{A_{m}}FF\left(m^{\prime}\rightarrow m^{\prime},\Gamma,\mathcal{O}\right).
\end{align}

Note that \(B_{2}\) in Eq.(\ref{eq:TSF}) can only be extracted if the fit has access to multiple sink times \(t\) as only varying the current time \(\tau\) cannot distinguish \(B_{0}\) from \(B_{2}\).

Since we have access to multiple smearings, we can also construct a
combined fit over smearing-dependent \(A_{m}\) and \(A_{m^{\prime}}\)
but a common \(m\) and \(\Delta m\).

The process for the two-exponential fit is to fit the two-point
correlator over a sink time range in which the two-state ansatz is
justified. Then use these extracted parameters in the fit to the
three-point correlator using a \(\tau\) range that also satisfies a
two-state ansatz.

Given the experience in spectroscopy studies, we emphasise that the
fit parameter $\Delta m$ should not be taken too literally in terms of
the energy gap to the first excited states. The exponential behaviour
is merely acting to mock up the sum of all excited states over the
range of fit considered. It is for this reason we prefer the
nomenclature ``two-exponential fit'' instead of ``two-state fit''.

\section{\label{Res}Results}
\subsection{\label{Res:TwoCor} Two Point Correlator}

\begin{figure}
\includegraphics[trim={4mm 5mm 4mm 3mm},clip,width=.48\textwidth]{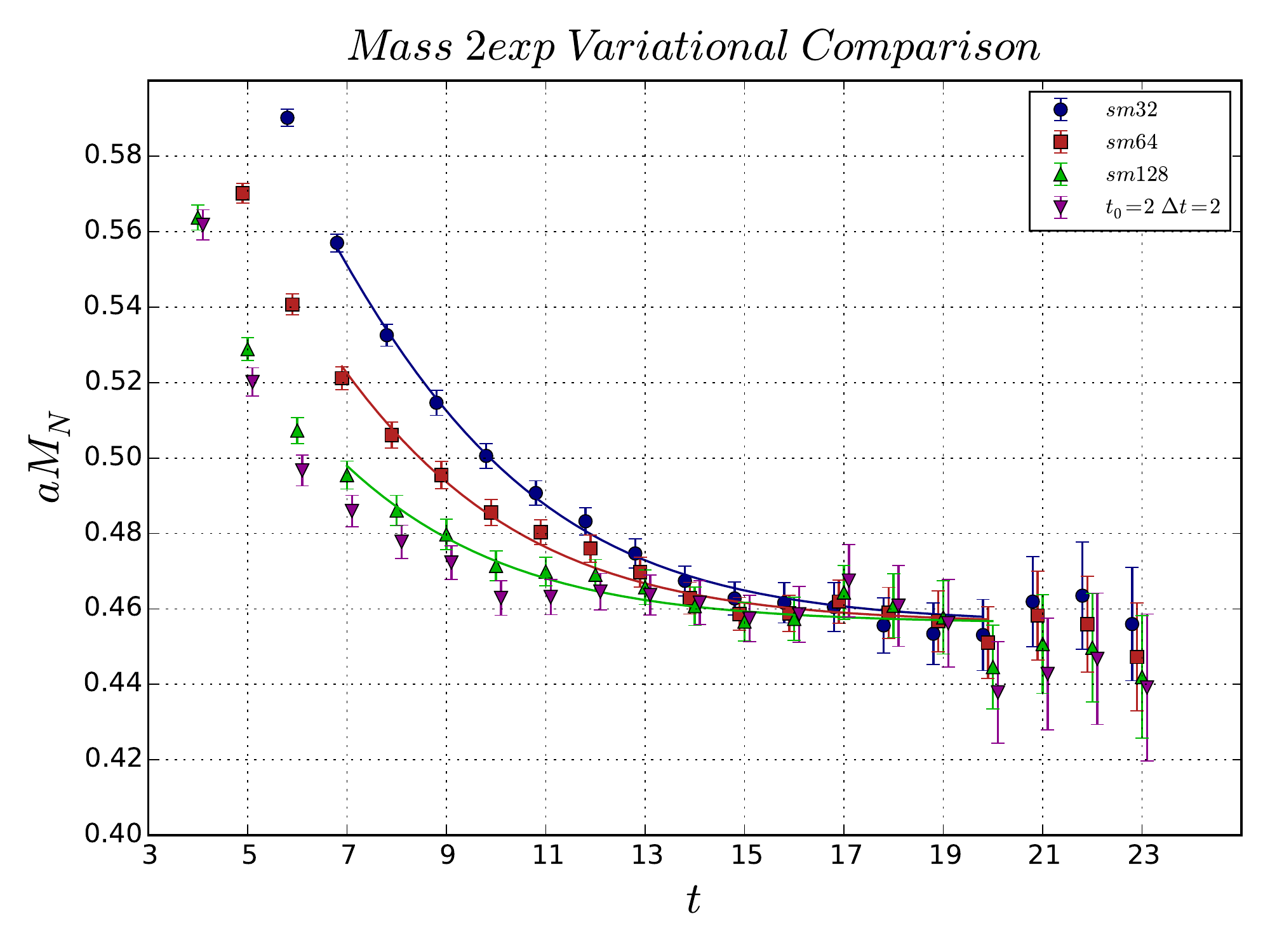}
\caption{\label{MassGraph}Mass plots over sink time (using \(\Delta t=3\) in Eq.(\ref{eq:MApx})) comparing the different smearings (circle, square, triangle) and the variational method (up side down triangle). The lines plotted are the two-exponential fit results described in Section \ref{TSF}.}
\end{figure}

Initial analysis is done on the two-point correlator for the variational analysis, since it is needed for the construction of the combined sink smearing. Via the standard construction below, we can extract the mass assuming a sufficiently large Euclidean time is taken.
\be \label{eq:MApx}
\log\left(\frac{\GtwoptGfour}{\Gtwoptpdt}\right) \xrightarrow{t \gg 0} \Ep^{0} \Delta t .
\ee
This function is also graphed for visualisation (Figure~\ref{MassGraph}) with the two-exponential fit function fitted to all source-sink smearing amounts along side the variational method.

By looking at the mass plots (Figure~\ref{MassGraph}) we can see that the variational method is producing a correlator similar to the 128 sweeps of smearing result, but with more excited states being removed. The two-exponential fit seems to indicate that the mass plateau is lower to where you might expect to get a good \(\chi^2_{PDF}\) for a single state fit in the variational method.

\subsection{\label{Res:Axial} Nucleon axial charge \(g_{A}\)}
The nucleon axial charge has been quite an important benchmark for the
validity of lattice QCD calculations. It can be calculated by looking
at the operator \(\mathcal{O}=i\gamma_{3}\gamma_{5}\) while using a
spin projector which corresponds to \(\Gamma = \Gamma_{3} \equiv
\left(\frac{I+\gamma_{4}}{2}\right)\gamma_{3}\gamma_{5}\).
%


\begin{figure}
\includegraphics[trim={4mm 5mm 4mm 3mm},clip,width=.48\textwidth]{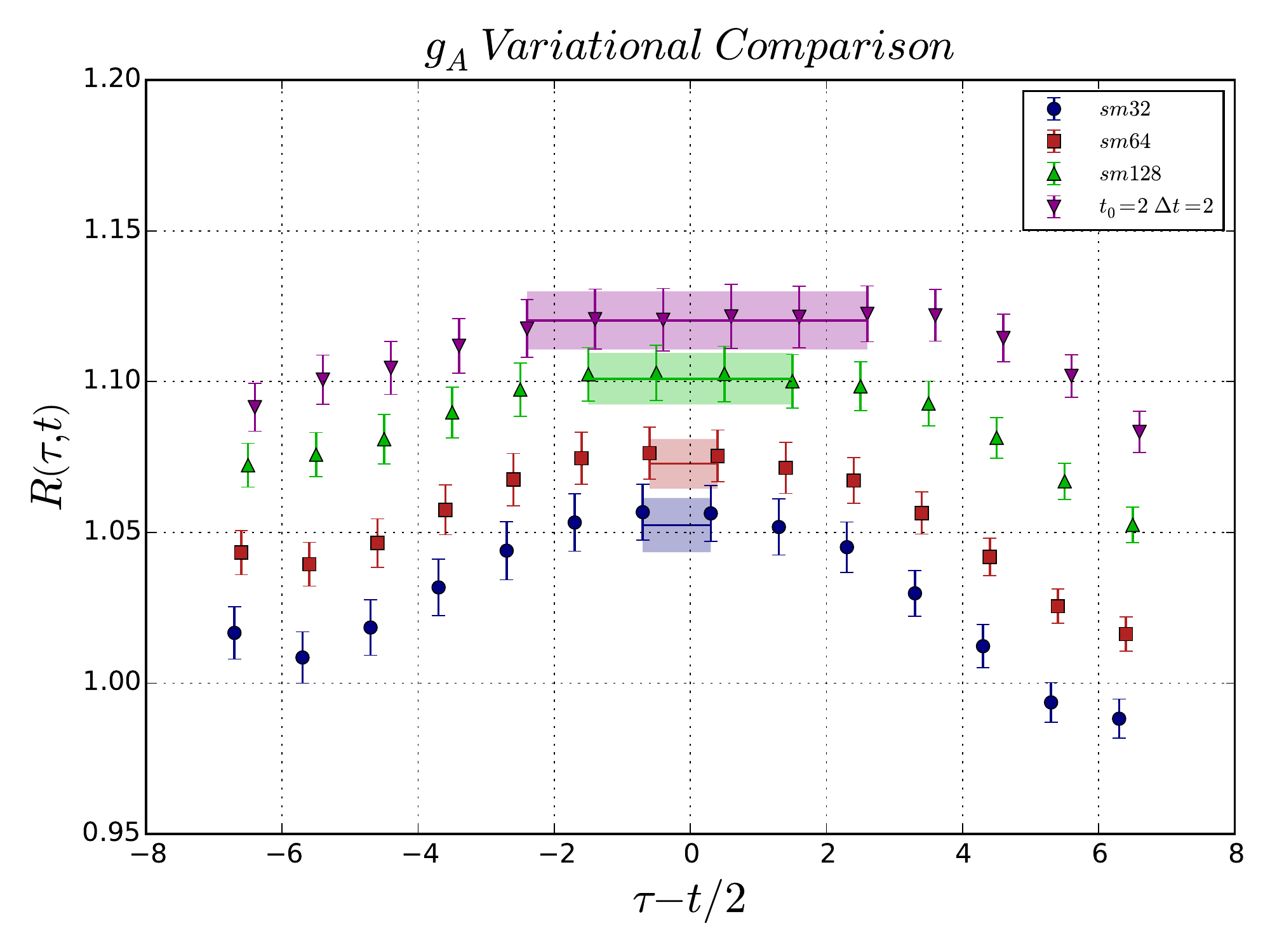}
\caption{\label{gAGraph}Graph for \(g_{A}\) extracted from the R function defined in Eq.(\ref{eq:Rfac}). The lines indicate the constant fit value extracted from each set of data used in Figure~\ref{gASummary}. All subsequent R function graphs are plotted over the current insertion time \(\tau\) which has been centred about zero via the construction \(\tau-t/2\). This plot compares different diagonally smeared values to the variational method. All results had a source-sink separation of \(t=13\).}
\end{figure}

In Figure~\ref{gAGraph}, we plot the ratio in Eq.(\ref{eq:Rfun}) over the current insertion time, using \(N_{smear} = 32, 64, 128\) at both source and sink, along with the variational method all at a fixed source-sink separation of 13. For the smeared results, we see that no clear plateau is present around the central current insertion point. In contrast, we can see that the variational method seems to have removed majority of the contamination from transition matrix elements as it looks to plateau from current insertion time 5 to 11. Furthermore, the value produced is statistically larger than any of the smeared results indication that a poor choice of source and sink operators and/or short source-sink separation times can lead to excited state contamination which acts to suppress \(g_{A}\). This is in agreement with other findings \cite{Bali:2014nma,Bali:2014nma,Yoon:2016dij}.

\begin{figure}
\includegraphics[trim={4mm 5mm 4mm 3mm},clip,width=.48\textwidth]{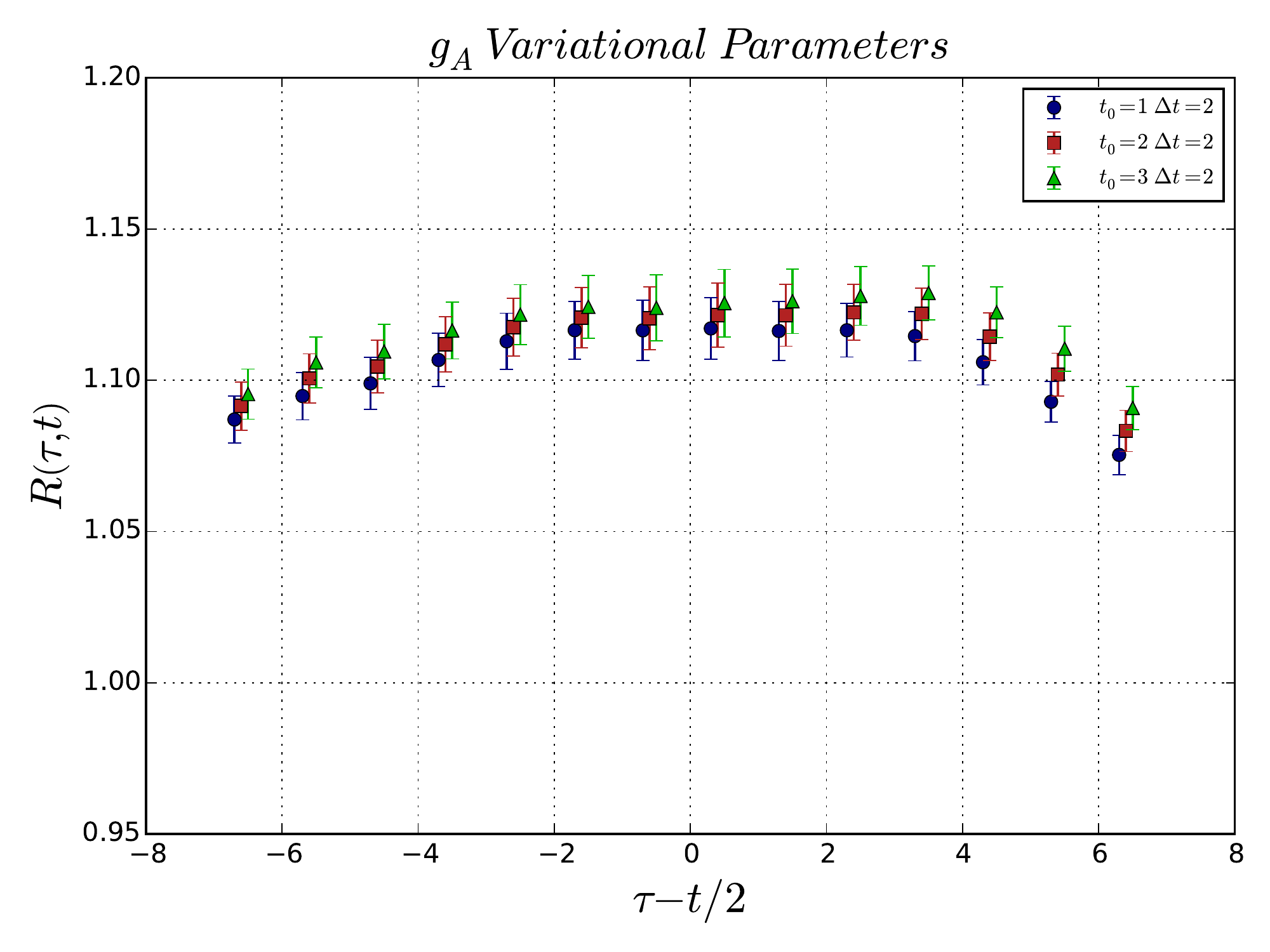}
\caption{\label{gAVarCompGraph}Graph for \(g_{A}\) extracted at each current time \(\tau\) comparing all the variational methods produced by varying the \(t_{0}\) and \(\Delta t\) parameters. This was calculated at a source-sink separation of \(t=13\).}
\end{figure}

Since we have access to the full 3x3 correlation matrix at a source-sink separation of 13, it is possible to utilise any \(t_{0}\) and \(\Delta t\) calculated in the two-point correlator case. Exploring these parameters in Figure~\ref{gAVarCompGraph} for the variational results with a source-sink separation of 13, we see that the variational method parameters \(t_{0}\) and \(\Delta t\) have minimal effect on the calculation. We choose \(t_{0}=2\) and \(\Delta t=2\) as it allowed sufficient time after the variational method diagonalisation for the correlator to reach the ground state.

\begin{figure}
\includegraphics[trim={4mm 5mm 4mm 3mm},clip,width=.48\textwidth]{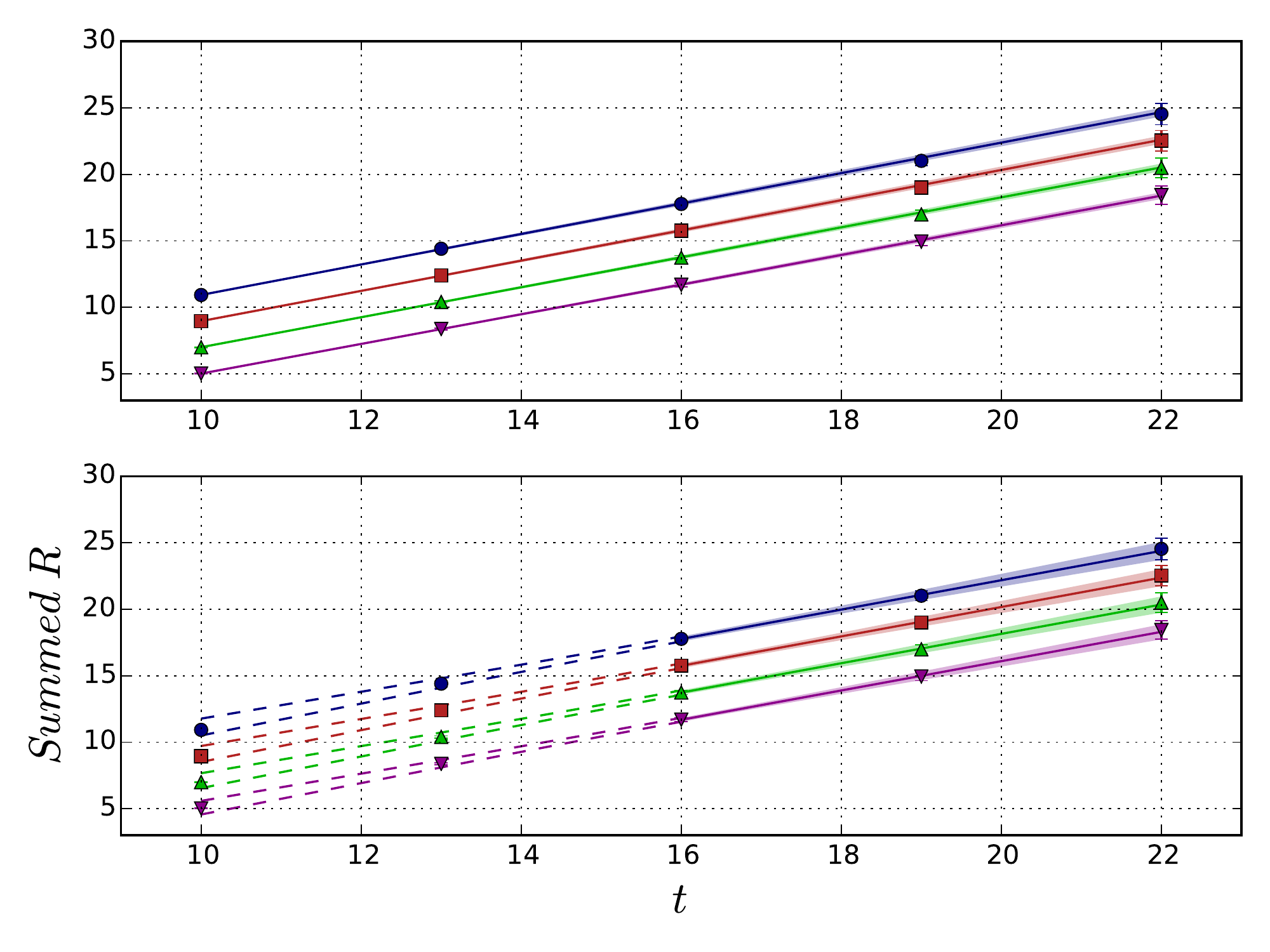}
\caption{\label{gASumGraph}Summed ratio factor values for multiple source-sink times defined in Eq.(\ref{eq:Sum}). The different colours/symbols (blue/circle, red/square, green/triangle and pink/up-side-down triangle respectively) correspond to \(\delta t = 0,1,2,3\) where \(\delta t\) is defined in in Eq.(\ref{eq:Sum}). The linear line of best fit is used to extract a slope which corresponds to \(g_{A}\) and the dashed lines correspond to the projected error ranges to smaller source-sink summed results.}
\end{figure}

In the plots in Figure~\ref{gASumGraph} we have the summation function defined in Eq.(\ref{eq:Sum}) for \(g_{A}\) plotted over the source-sink separation times (in which we have summed over the current insertion times). The colours/symbols blue/circle, red/square, green/triangle and pink/up-side-down triangle let us see the change in the line of best fit when we vary \(\delta t=0,1,2,3\) respectively in Eq.(\ref{eq:Sum}). The top plot shows that the summation fits show no statistically significant change in slope for the different \(\delta t\) value results and the line of best fit seems to satisfy the points well to extract a value. Results with small source-sink separations are likely to have the most contamination from higher excitations. They also have smallest statistical error and so can dominate in a weighted fit. By fitting only to the largest 3 source-sink separated results, we can extend the lines back to compare with the smaller source-sink separated results. Any significant deviation indicates that those smaller source-sink separated results should be excluded from the final fit. For \(g_{A}\) in the bottom plot in Figure~\ref{gASumGraph}, we have excluded the two smallest source-sink separated points from the linear fit and we see that the projected errors do encapsulate the smaller source-sink separated results. We can also see that the errors on the results drastically increase when compared to the top figure, but we see no more \(\delta t\) dependence which is required if we are to accept the first order transitional matrix element approximation.



%

\begin{figure}
\includegraphics[trim={4mm 5mm 4mm 3mm},clip,width=.48\textwidth]{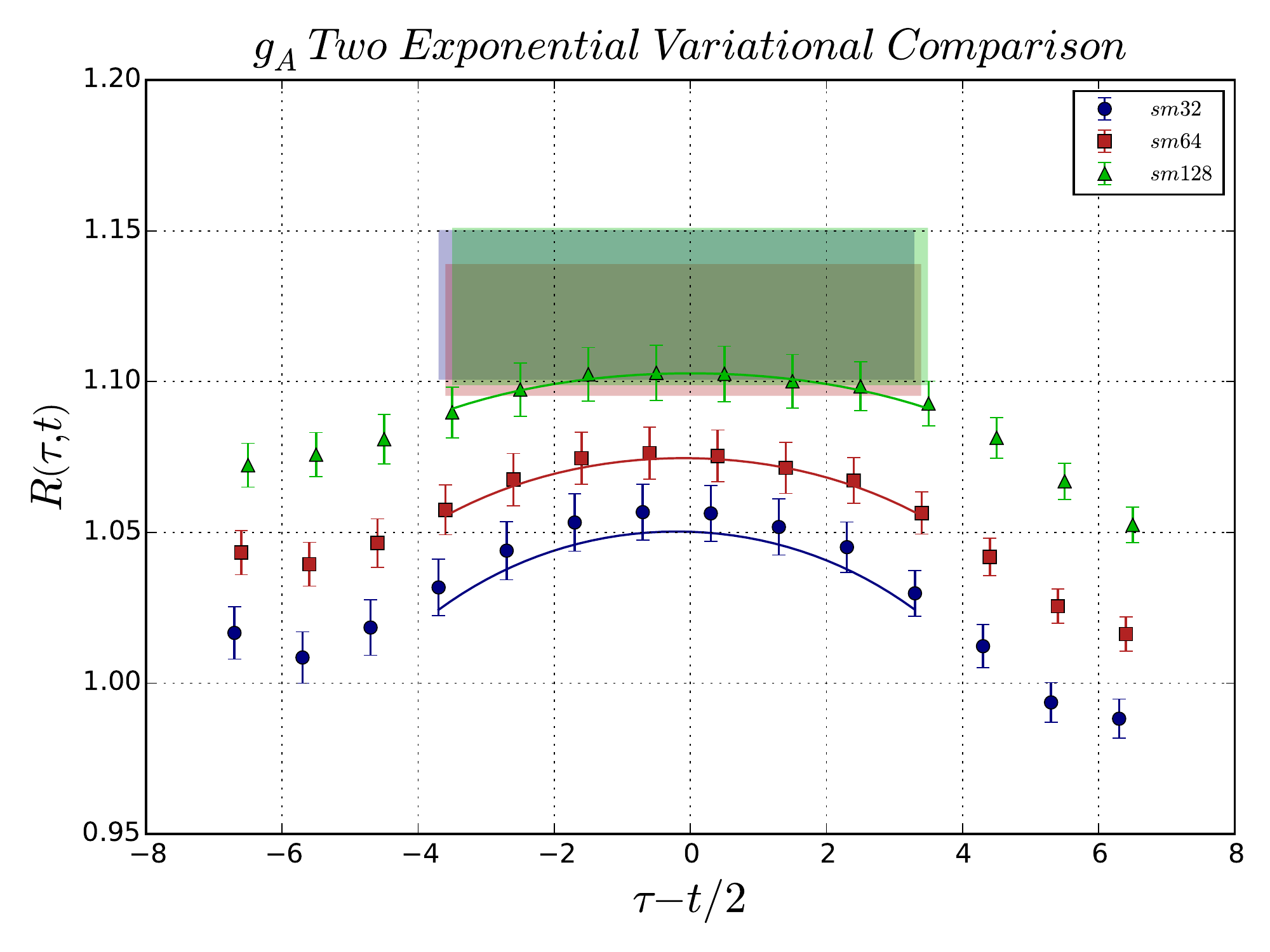}
\caption{\label{gA2expVarGraph}Graph of Figure~\ref{gAGraph} overlaid with a two-exponential fit calculation on each set. The lines are the corresponding fit function and the shaded areas corresponds to \(g_{A}\) value extracted from the fit parameters. Since there is no varying in sink time \(t\), no \(B_{2}\) was calculated in Eq.(\ref{eq:TSF}) as discussed in the end of Section \ref{TSF}.}
\end{figure}

In applying the two-exponential fit to the differently smeared results at a source-sink separation of 13
in Figure~\ref{gA2expVarGraph} (for \(g_{A}\)), all three smearing fits coincided with one another, having a larger relative error compared to the data points fitted to and being statistically consistent with a constant fit to the largest smeared (sm128) result.

\begin{figure}
\includegraphics[trim={4mm 5mm 4mm 3mm},clip,width=.48\textwidth]{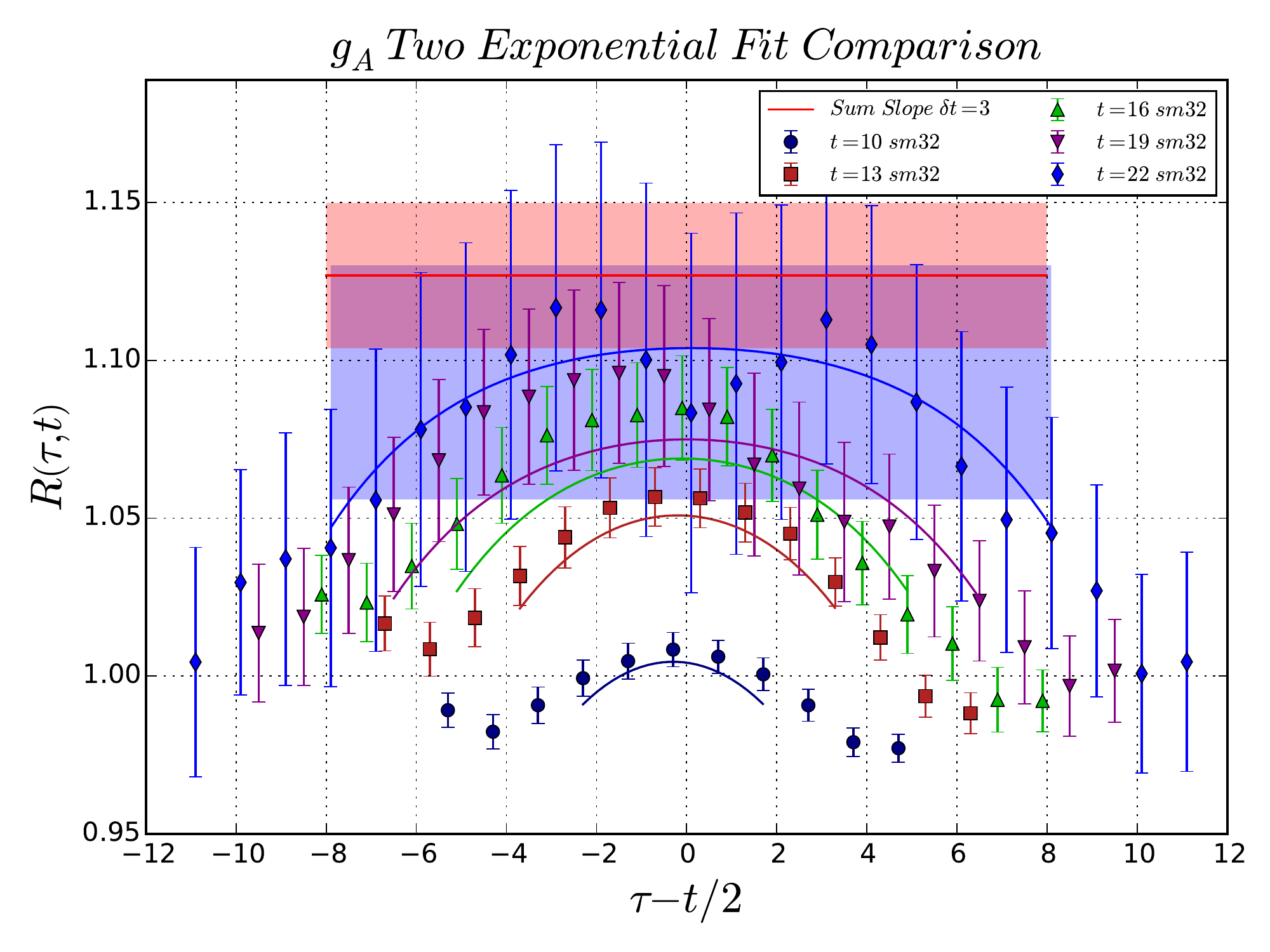}
\caption{\label{gA2expTsinkGraph}Graph for \(g_{A}\) extracted at each current time \(\tau\) for multiple sink times plotted as a comparison to the summation method using a \(\delta t = 3\) (purple line in Figure~\ref{gASumGraph}). Also overlaid with a single two-exponential fit over both current and sink time (\(t\) and \(\tau\)). The lines correspond to the two-exponential fit function constructed and the blue shaded area corresponds to the \(g_{A}\) parameter extracted from the two-exponential fit. All results are calculated with 32 multiples of smearing at the source and sink.}
\end{figure}

For \(g_{A}\), doing a combined fit to all the source-sink separated data as in Figure~\ref{gA2expTsinkGraph} leads to a result that is very similar to a constant fit for the largest source-sink separated result. Similar to the summation method, the two-exponential method is heavily weighted by the smallest source-sink separated values which can be problematic as these values are most susceptible to excited state contamination.

\begin{figure}
\includegraphics[trim={4mm 5mm 4mm 3mm},clip,width=.48\textwidth]{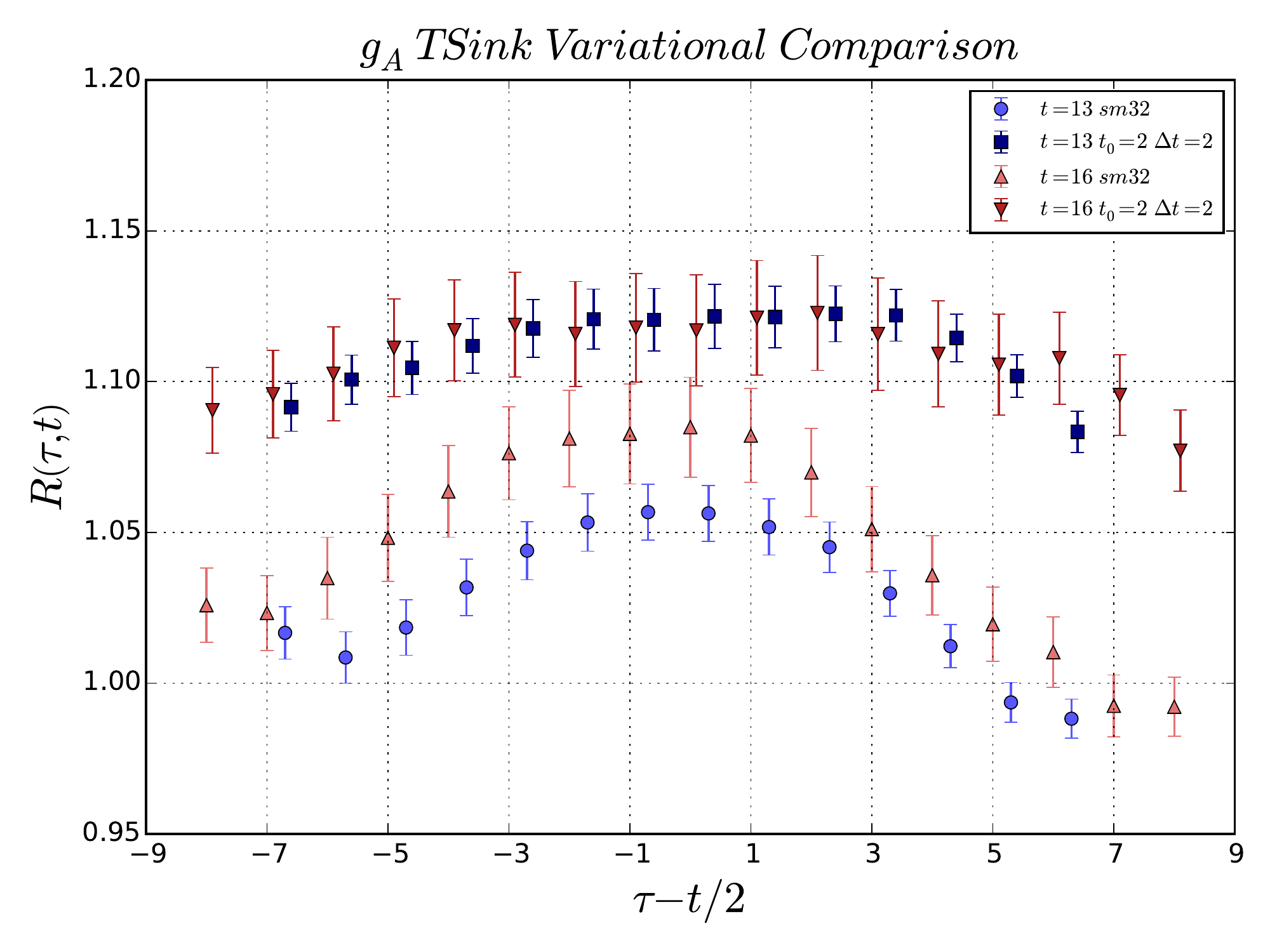}
\caption{\label{gATSinkVarGraph}Graph for \(g_{A}\) extracted at each current time \(\tau\) comparing different variational results for different source-sink separations (square and up-side-down triangle points). This is overlaid with lighter coloured results which is the corresponding source-sink separation for the 32 sweeps of smearing result (circle and triangle points).}
\end{figure}

Figure~\ref{gATSinkVarGraph} shows that for the variational method calculation for \(g_{A}\), there are no more excited states to remove as the results did not shift up when moving from a source-sink separation of 13 to 16. Compared to the smallest smeared operators, we see excited states being removed in the change from a source-sink separation of 13 to 16.

\begin{figure}
\includegraphics[trim={4mm 5mm 4mm 3mm},clip,width=.48\textwidth]{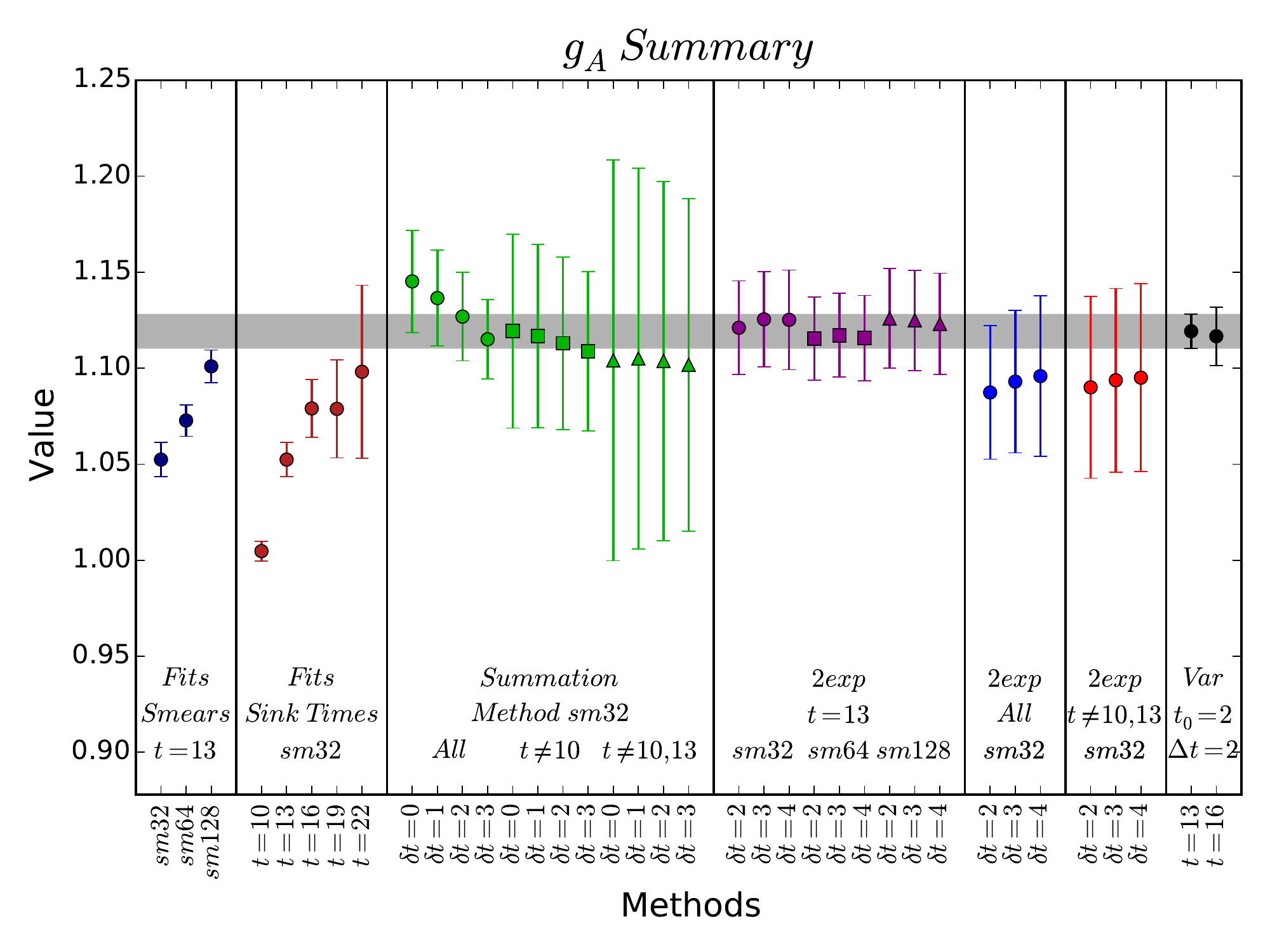}
\caption{\label{gASummary}Summary of all the extracted values for \(g_{A}\) over the different methods. ``Fits'' and ``Var'' use a constant fit range to extract a value. \(\delta t\) is defined in Eq.(\ref{eq:Sum}) for the summation method and corresponds to how many points are excluded from the source and sink current times in two-exponential fit range. ``All'' refers to using all the source-sink separated results in the analysis and \(t\ne n\) refers to excluding \(n\) in the calculation. The shaded band helps to compare the variational method \(t=13\) result.}
\end{figure}

In the final summary plot for \(g_{A}\) containing all the extracted values from all the different methods calculated (Figure~\ref{gASummary}), we see that the variational method demonstrates reliability and robustness as it produces a value that improves on the results that alter the smearing amounts and small source-sink separated results by removing excited states and improves on the summation and two-exponential fit method by producing a much more precise result. The variational method result of \(g^{Var}_{A}=1.1203(96)\) agrees within statistical error with the Feynman-Hellmann theorem result of \(g^{FH}_{A}=1.101(24)\) \cite{Chambers:2014qaa} on the same set of gauge field configurations that are used in this work.

\subsection{\label{Res:Icur} Scalar Current}

The scalar current form factor has been notorious for its large excited state contamination. It can be calculated by looking at the operator \(\mathcal{O}=I\) while using a spin projector \(\Gamma = \Gamma_{4} \equiv \frac{I+\gamma_{4}}{2}\) which corresponds to an unpolarised nucleon. The same analysis can be undertaken for this operator at zero source and sink momentum which leads to a result for the isovector scalar charge, \(g_{S}\).

\begin{figure}
\includegraphics[trim={4mm 5mm 4mm 3mm},clip,width=.48\textwidth]{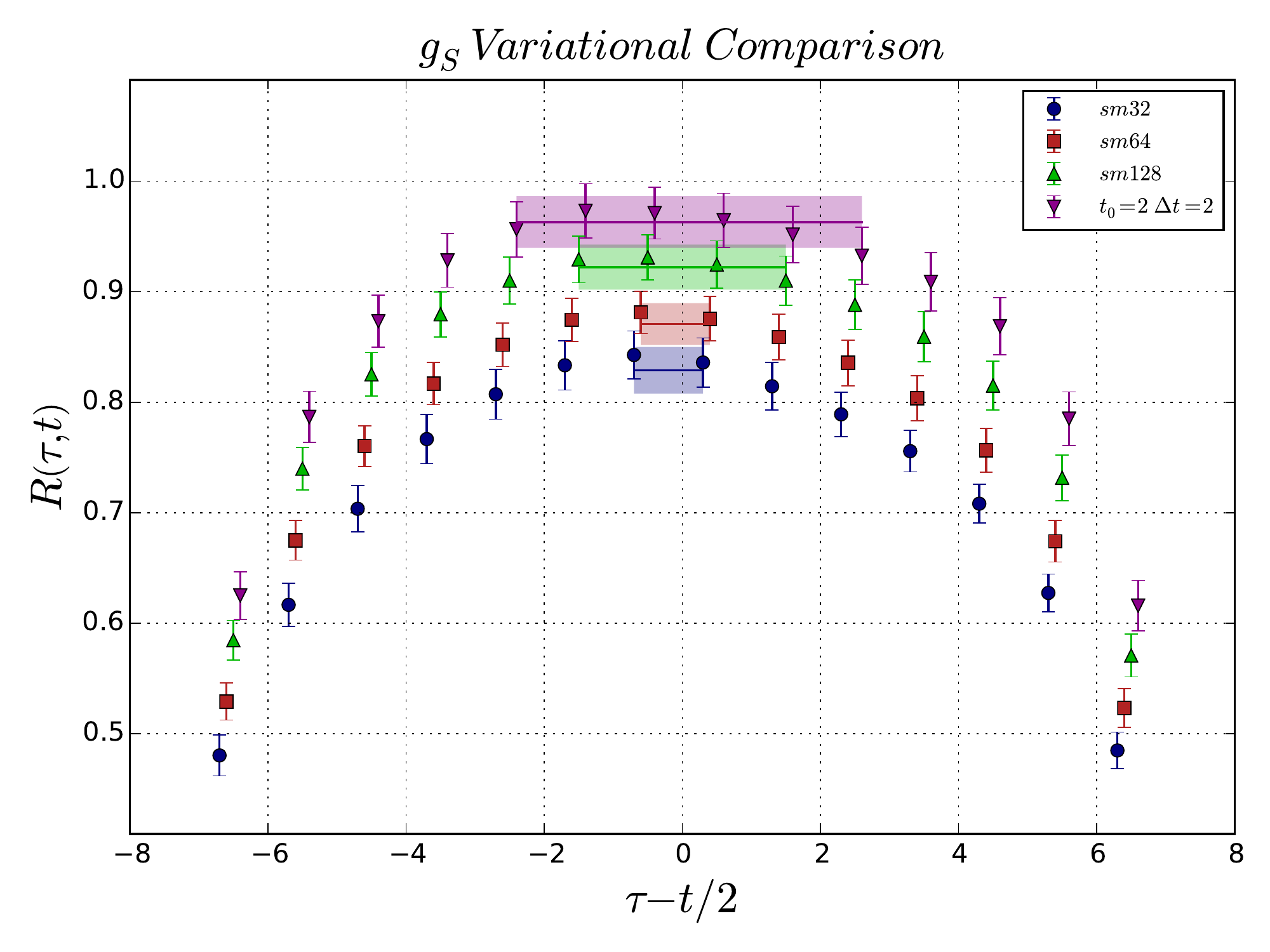}
\caption{\label{IGraph}Graph for \(g_{S}\) extracted at each current time \(\tau\), comparing different diagonally smeared values to the variational method. The lines indicate the constant fit value extracted from each set of data used in Figure~\ref{ISummary}. This was calculated at a source-sink separation of \(t=13\).}
\end{figure}

In Figure~\ref{IGraph}, we see for \(g_{S}\) the variational method producing a flatter ratio as a function of \(\tau\) compared to the individually smeared correlators. We note that in this case, we see that the transition matrix elements are much larger than \(g_{A}\) as there is a larger curvature with respect to current time insertion \(\tau\).

\begin{figure}
\includegraphics[trim={4mm 5mm 4mm 3mm},clip,width=.48\textwidth]{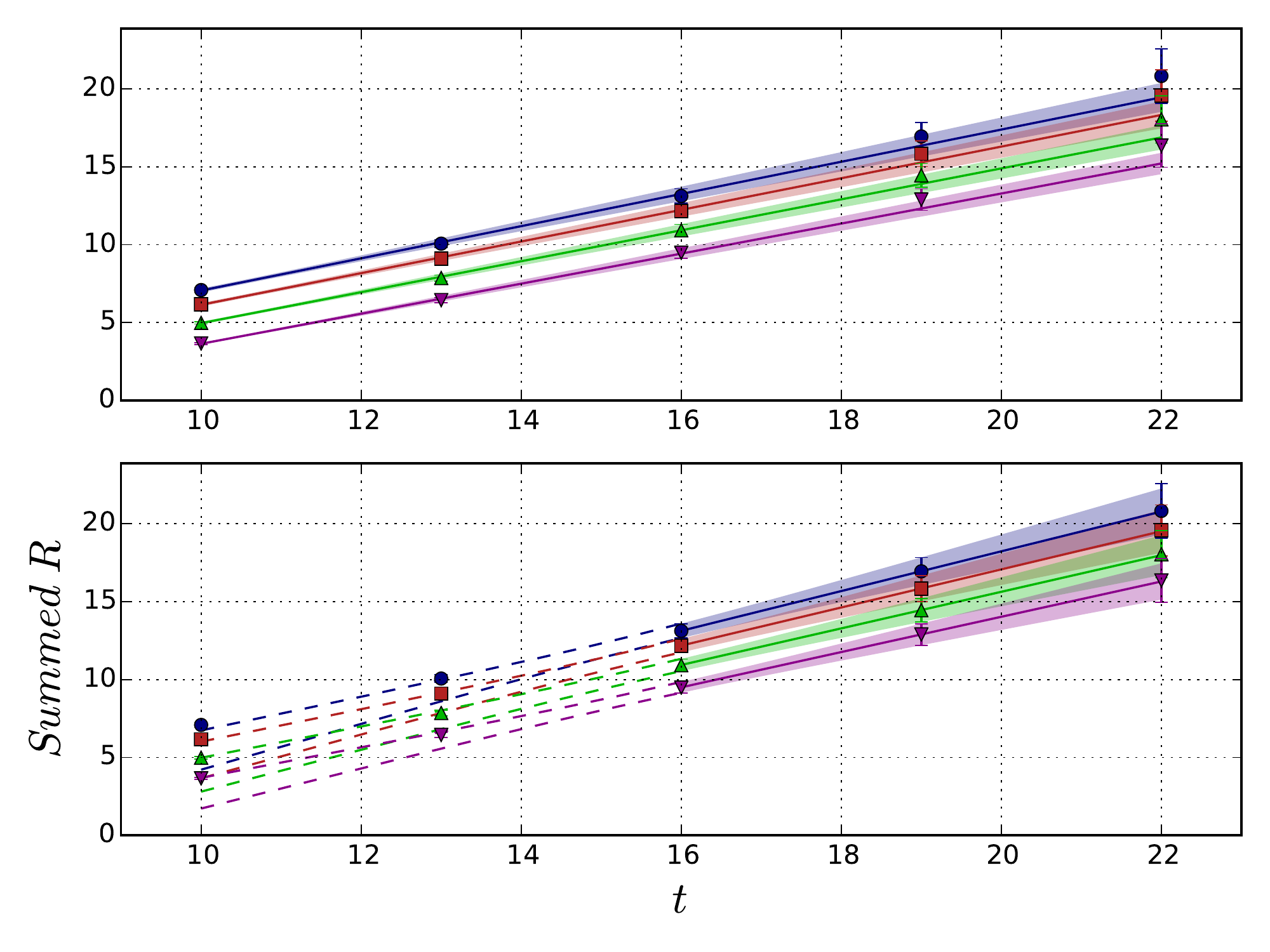}
\caption{\label{ISumGraph}Summed ratio factor values for multiple source-sink times defined in Eq.(\ref{eq:Sum}). The different colours/symbols (blue/circle, red/square, green/triangle and pink/up-side-down triangle respectively) correspond to \(\delta t = 0,1,2,3\) where \(\delta t\) is defined in in Eq.(\ref{eq:Sum}). The linear line of best fit is used to extract a slope which corresponds to \(g_{S}\) and the dashed lines correspond to the projected error ranges to smaller source-sink summed results.}
\end{figure}

In the summation method results, comparing the 4 coloured slopes passing through the 4 colours/symbols (blue/circle, red/square, green/triangle and pink/up-side-down triangle respectively) in the top of Figure~\ref{ISumGraph} shows that the \(\delta t = 0,1,2,3\) parameter variation is not statistically significant. However, as the fit is a weighted fit and the smallest source-sink separated points have the smallest errors and the set of points are not linear, the smallest points are forcing the linear function to underestimate the slope of the larger source-sink separated values. Fitting over the larger source-sink separated points in the bottom of Figure~\ref{ISumGraph} and projecting the fit backwards to smaller times reveals a tension between the results at small and large source-sink separations as the projected errors do not encapsulate the smaller source-sink separated results. This suggests that the error term in Eq.(\ref{eq:Sum}) is starting to be statistically significant.

%
%

%

\begin{figure}
\includegraphics[trim={4mm 5mm 4mm 3mm},clip,width=.48\textwidth]{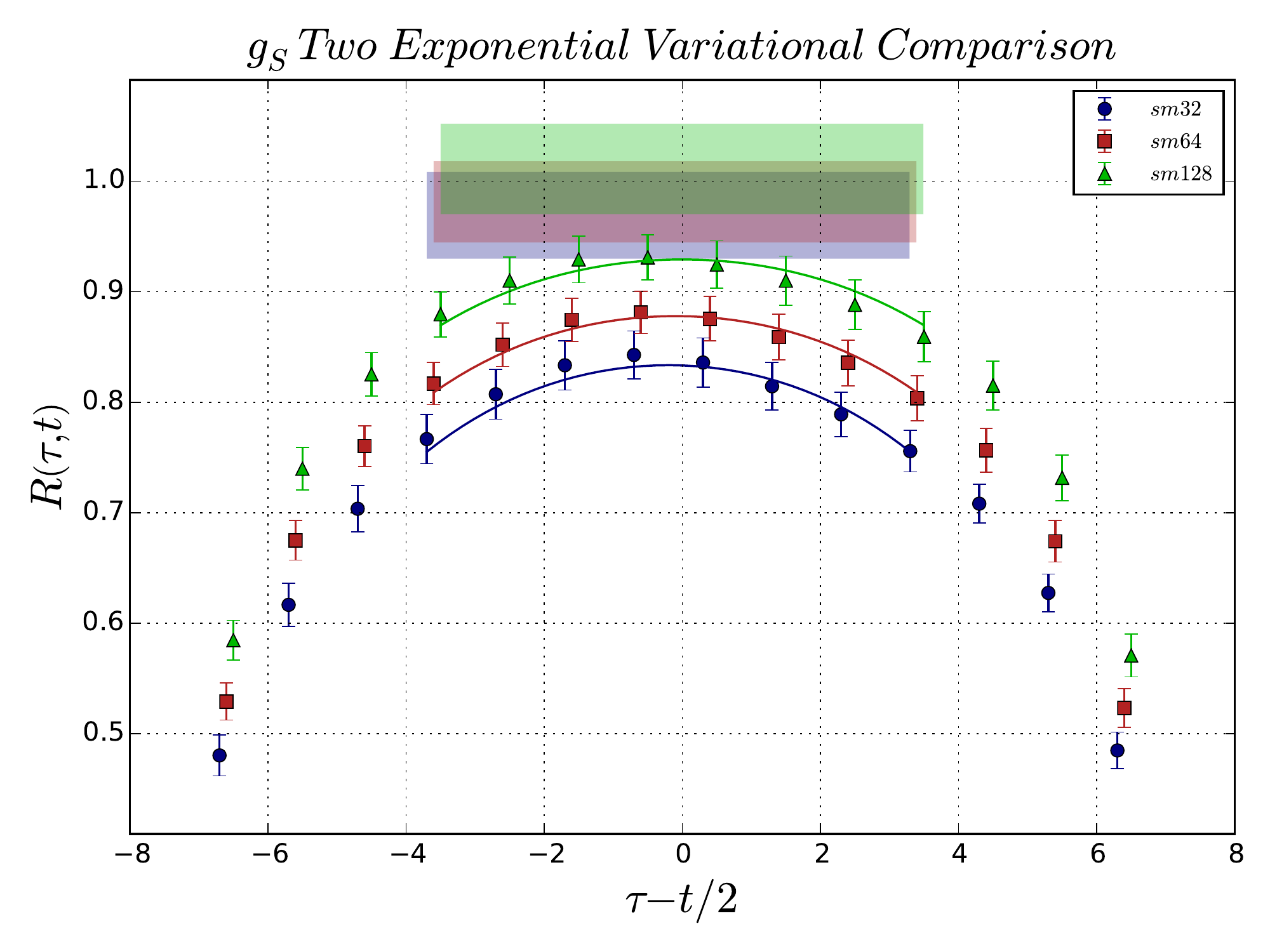}
\caption{\label{I2expVarGraph}Graph of Figure~\ref{IGraph} overlaid with a two-exponential fit calculation on each set. The lines are the corresponding fit function and the shaded areas corresponds to \(g_{S}\) value extracted from the fit parameters. Since there is no varying in sink time \(t\), no \(B_{2}\) was calculated in Eq.(\ref{eq:TSF}) as discussed in the end of Section \ref{TSF}.}
\end{figure}

Applying the two-exponential fit to \(g_{S}\) for the smeared results in Figure~\ref{I2expVarGraph}, appears to have made an improvement to all 3 smeared results. The errors on the parameter extracted has increased compared to the errors associated with the current insertion \(\tau\) points.

\begin{figure}
\includegraphics[trim={4mm 5mm 4mm 3mm},clip,width=.48\textwidth]{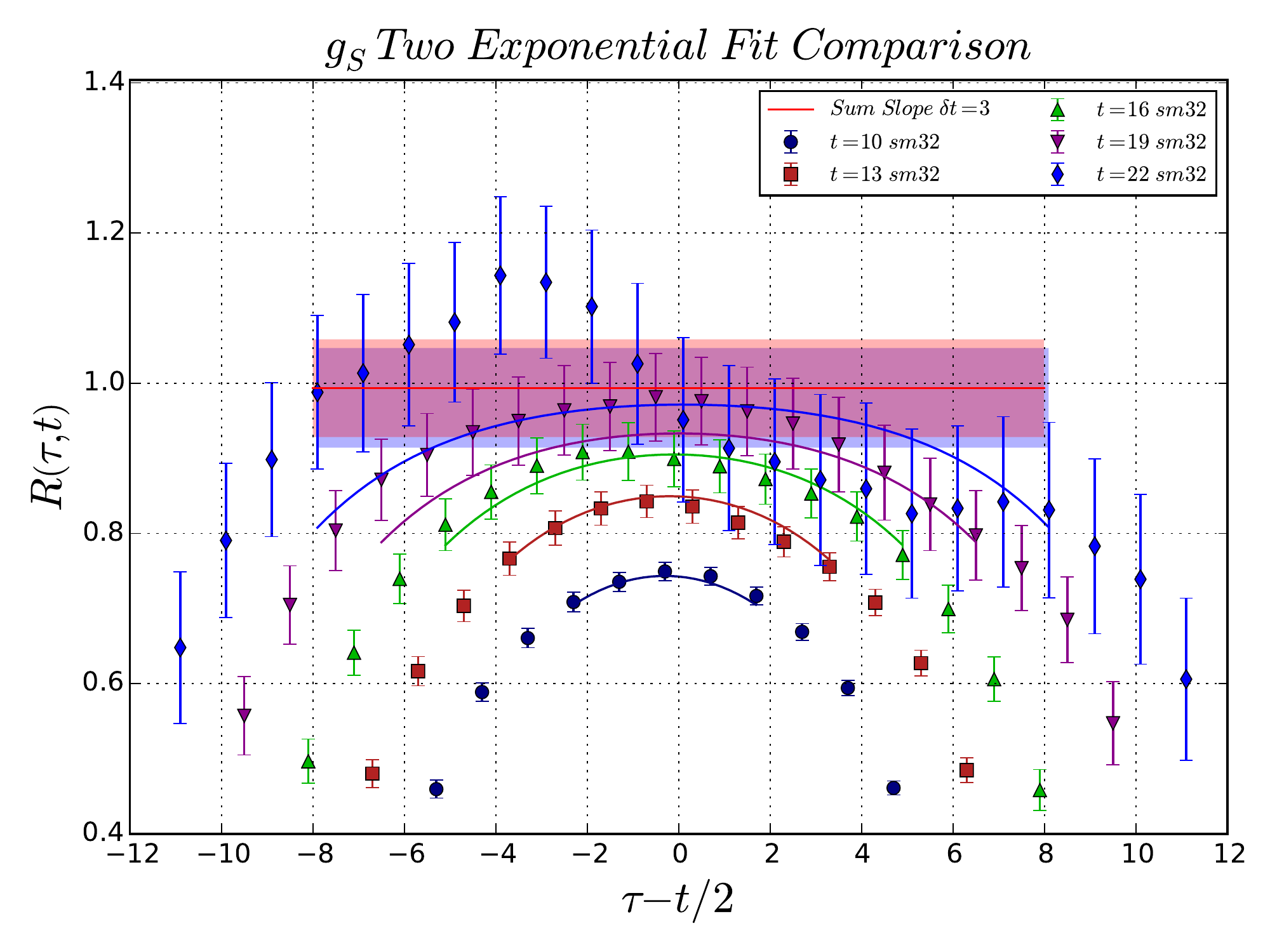}
\caption{\label{I2expTsinkGraph}Graph for \(g_{S}\) extracted at each current time \(\tau\) for multiple sink times plotted as a comparison to the summation method using a \(\delta t = 3\) (purple line in Figure~\ref{ISumGraph}). Also overlaid with a single two-exponential fit over both current and sink time (\(t\) and \(\tau\)). The lines correspond to the two-exponential fit function constructed and the blue shaded area corresponds to the \(g_{S}\) parameter extracted from the two-exponential fit. All results are calculated with 32 multiples of smearing at the source and sink.}
\end{figure}

The two-exponential fit to \(g_{S}\) in Figure~\ref{I2expTsinkGraph} again raises a lot of concern over the inclusion of small source-sink separations into the fit. Since the fit is weighted heavily to the smaller source-sink separated results, due to their statistical error the larger source-sink separated results are almost ignored.

\begin{figure}
\includegraphics[trim={4mm 5mm 4mm 3mm},clip,width=.48\textwidth]{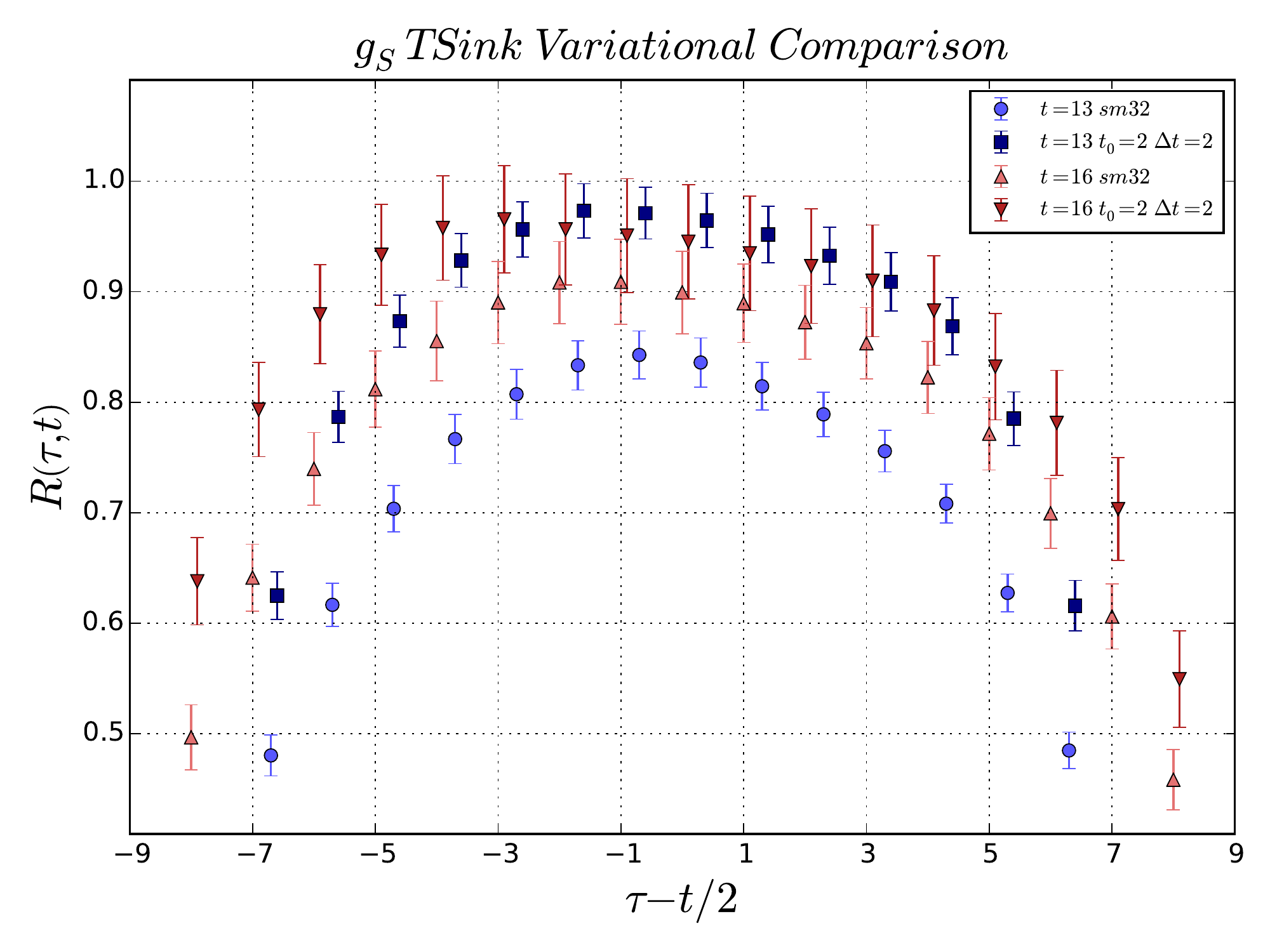}
\caption{\label{ITSinkVarGraph}Graph for \(g_{S}\) extracted at each current time \(\tau\) comparing different variational results for different source-sink separations (square and up-side-down triangle points). This is overlaid with lighter coloured results which is the corresponding source-sink separation for the 32 sweeps of smearing result (circle and triangle points).}
\end{figure}

Once again for \(g_{S}\) in Figure~\ref{ITSinkVarGraph}, increasing the source-sink separation for the variational method shows no more statistically significant removal of excited states which cannot be said about the smallest smeared result.

\begin{figure}
\includegraphics[trim={4mm 5mm 4mm 3mm},clip,width=.48\textwidth]{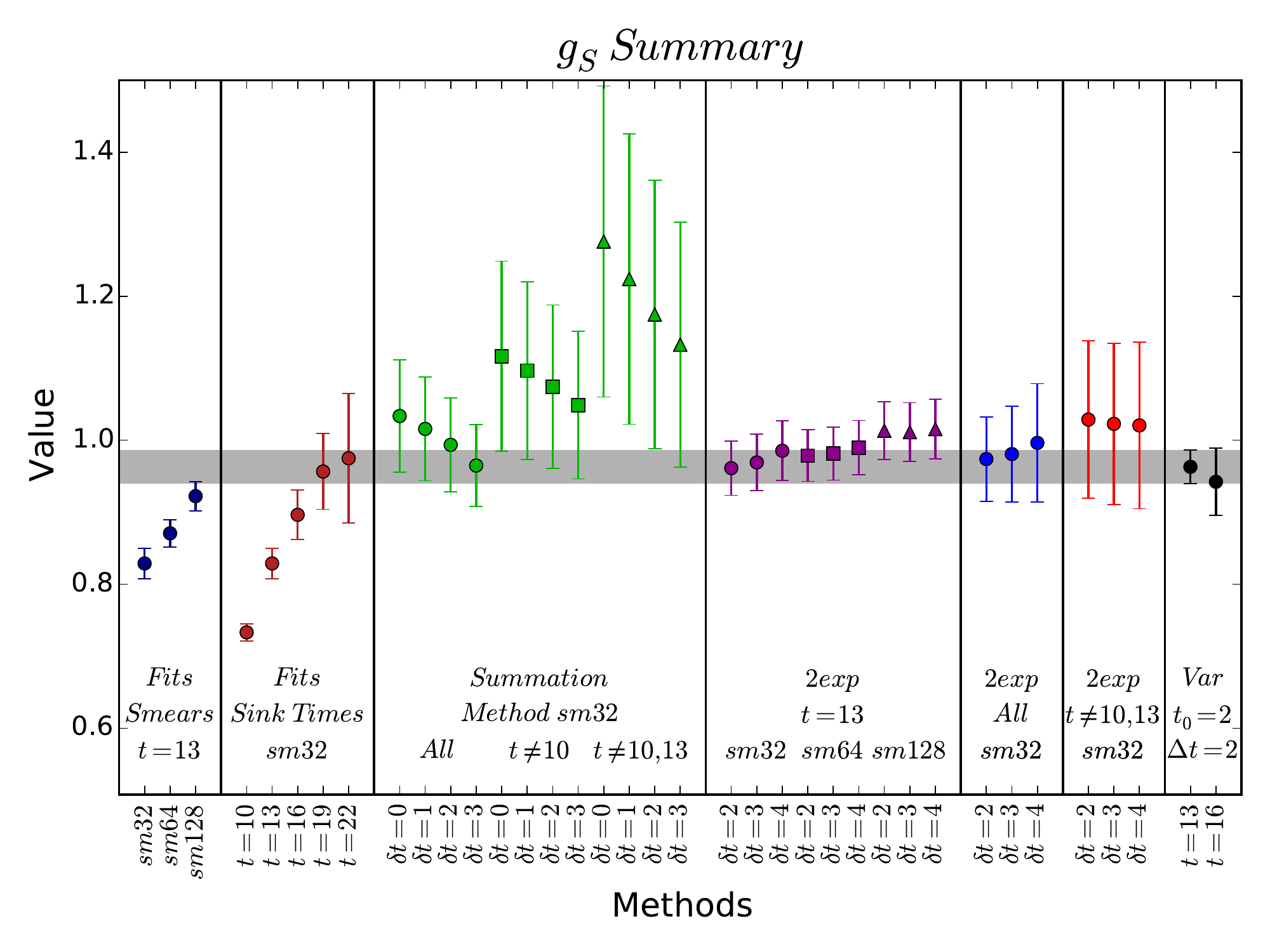}
\caption{\label{ISummary}Summary of all the extracted values for \(g_{S}\) over the different methods. ``Fits'' and ``Var'' use a constant fit range to extract a value. \(\delta t\) is defined in Eq.(\ref{eq:Sum}) for the summation method and corresponds to how many points are excluded from the source and sink current times in the two-exponential fit range. ``All'' refers to using all the source-sink separated results in the analysis and \(t\ne n\) refers to excluding n in the calculation. The shaded band helps to compare the variational method \(t=13\) result.}
\end{figure}

Similarly for the summary for \(g_{A}\), \(g_{S}\) in summary (Figure~\ref{ISummary}) shows that the variational method has removed all excited states and is a far more precise results compared to the summation and two-exponential fit methods. In addition, while not statistically significant, we observe an undesired \(\delta t\) dependence for each of the summation method results.

\subsection{\label{Res:MomFrac} Quark Momentum Fraction, \(\momfrac \)}
Deep inelastic scattering experiments are our primary method for understanding the nucleon and QCD in general. Looking at the operator product expansion, the momentum fraction carried by the quarks and gluons in the nucleon are directly related to the first moment of the structure functions. In any scheme and at any scale, the  quark and gluon momentum fractions sum to unity, providing good motivation for lattice QCD studies.

At the physical quark mass, it is predicted that \(\momfrac^{MRST} =0.157(9)\)\cite{Detmold:2001jb} where as the lattice determination of \(\momfrac \) at many quark masses has consistently over estimated the quantity over the years. One possible explanation could be due to the contamination from excited state effecting the results.

\(\momfrac \) can be calculated by looking at the operator
\(\mathcal{O}=\gamma_{4}D_{4}-\frac{1}{3}\left(\gamma_{1}D_{1}+\gamma_{2}D_{2}+\gamma_{3}D_{3}\right)
\) while using a spin projector which corresponds to \(\Gamma =
\Gamma_{4} \) as defined in the scalar current results section. The
same analysis can be undertaken for this combination. Note that the results presented here are for \(R(\tau,t)\) and haven't been converted to \(\momfrac\) or renormalised.

\begin{figure}
\includegraphics[trim={4mm 5mm 4mm 3mm},clip,width=.48\textwidth]{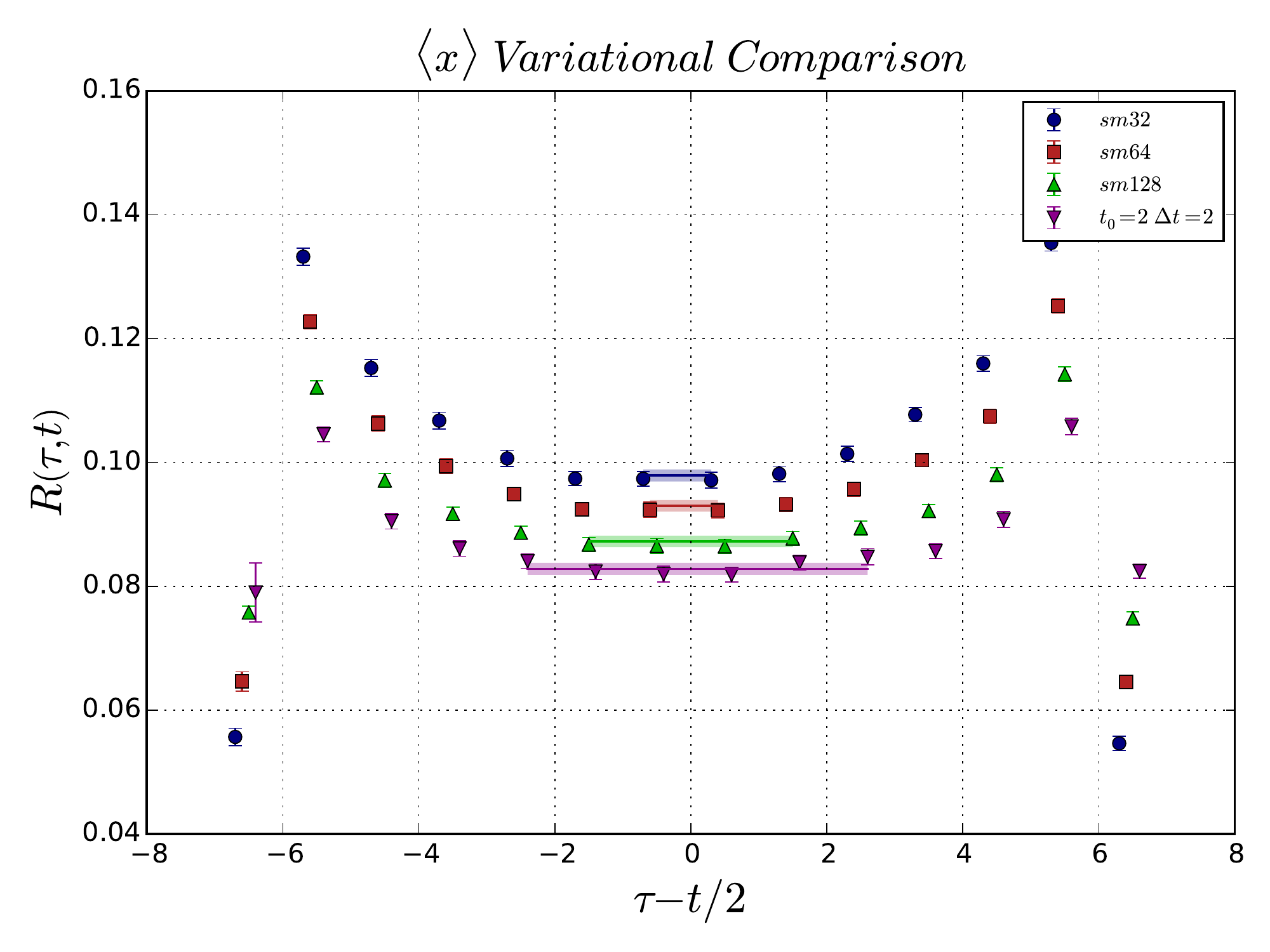}
\caption{\label{giDiGraph}Graph for \(R(\tau,t)\) using the operator for \(\momfrac\) extracted at each current time \(\tau\), comparing different diagonally smeared values to the variational method. This was calculated at a source-sink separation of \(t=13\).}
\end{figure}

A similar improvement as observed in the previous two quantities has been achieved by the variational method for \(\momfrac\) shown in Figure~\ref{giDiGraph}. For this operator we see there is much greater excited state contamination compared to the precision of the calculation of the current insertion time \(\tau\).

\begin{figure}
\includegraphics[trim={4mm 5mm 4mm 3mm},clip,width=.48\textwidth]{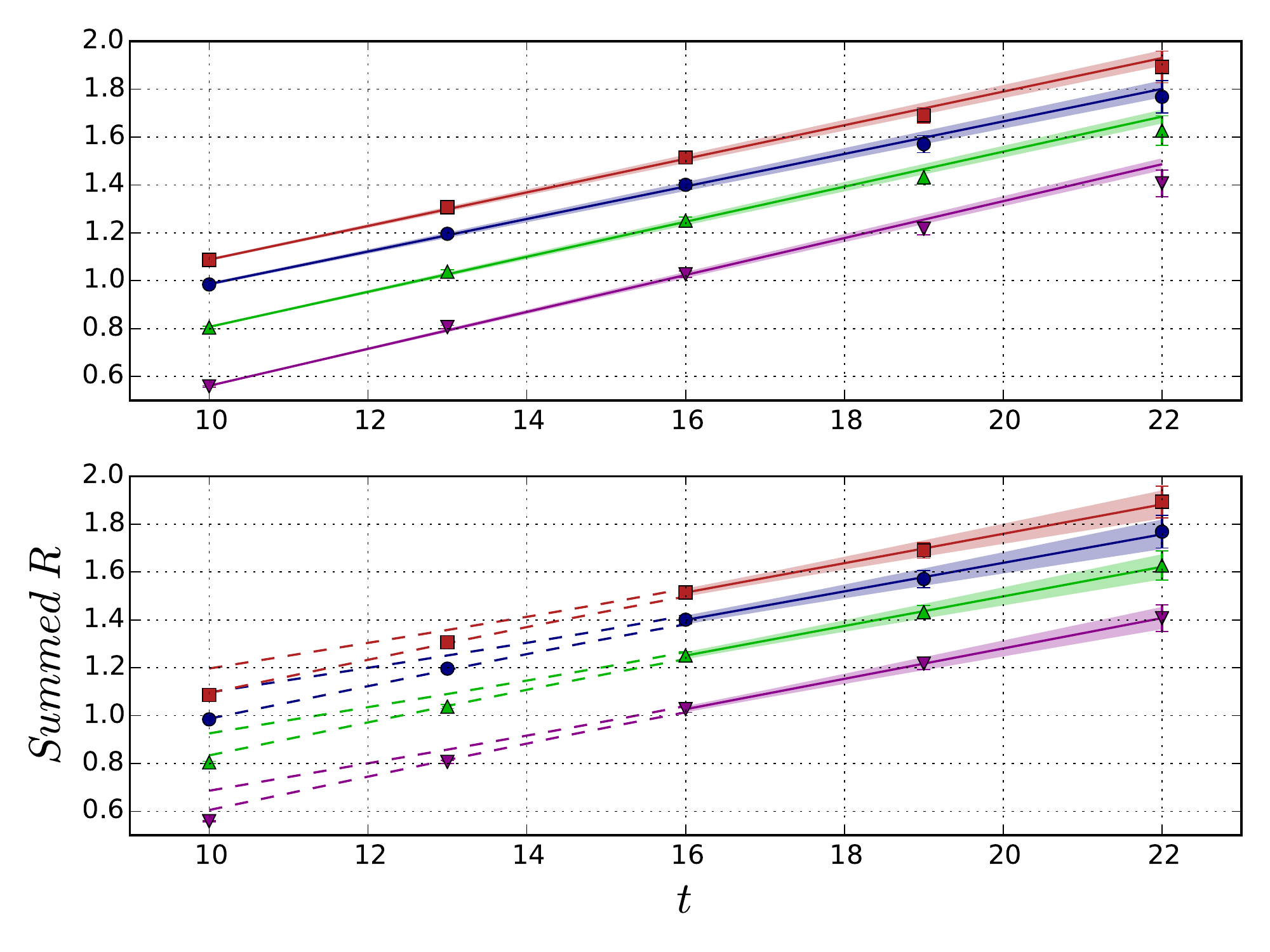}
\caption{\label{giDiSumGraph}Summed ratio factor values for multiple source-sink times defined in Eq.(\ref{eq:Sum}). The different colours/symbols (blue/circle, red/square, green/triangle and pink/up-side-down triangle respectively) correspond to \(\delta t = 0,1,2,3\) where \(\delta t\) is defined in in Eq.(\ref{eq:Sum}). The linear line of best fit is used to extract a slope which corresponds to \(\momfrac\) and the dashed lines correspond to the projected error ranges to smaller source-sink summed results.}
\end{figure}

Now the summation method fit undertaken in the top of Figure~\ref{giDiSumGraph} for \(\momfrac\) does show a variation on the \(\delta t\) parameter that is statistically significant. We can see for \(\momfrac\) the linear fit function is not sufficient to approximate the summed R function values. Again, fitting over larger source-sink separated points in the bottom of Figure~\ref{giDiSumGraph} and projecting the errors to smaller times shows that there is an inconsistency as the smaller source-sink separated result do not lie within the fit errors projected to smaller times. This tells us that the two-exponential approximation used in the summation method has broken down.

%

%

\begin{figure}
\includegraphics[trim={4mm 5mm 4mm 3mm},clip,width=.48\textwidth]{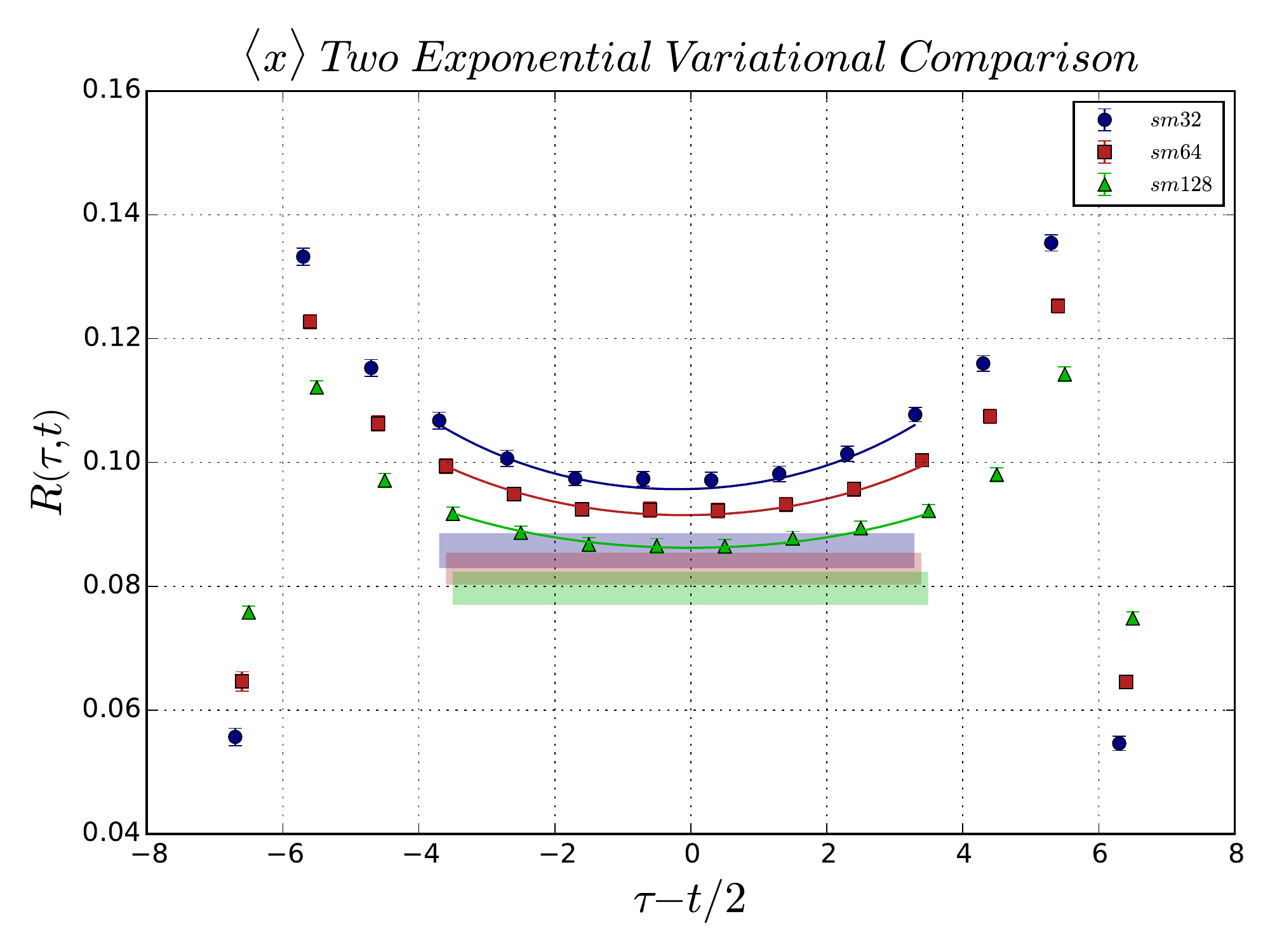}
\caption{\label{giDi2expVarGraph}Graph of Figure~\ref{giDiGraph} overlaid with a two-exponential fit calculation on each set. The lines are the corresponding fit function and the shaded areas corresponds to \(\momfrac\) value extracted from the fit parameters. Since there is no varying in sink time \(t\), no \(B_{2}\) was calculated in Eq.(\ref{eq:TSF}) as discussed in the end of Section \ref{TSF}.}
\end{figure}

Applying the two-exponential fit to \(\momfrac\) for the smeared results in Figure~\ref{giDi2expVarGraph}, it looks to have made an improvement to all 3 smeared results. The errors on the parameter extracted has increased compared to the error from a ratio function points, but for \(\momfrac\) it seems that the two-exponential fit was more successful due to the relative size of the excited state contamination to the precision of the ratio function points.

\begin{figure}
\includegraphics[trim={4mm 5mm 4mm 3mm},clip,width=.48\textwidth]{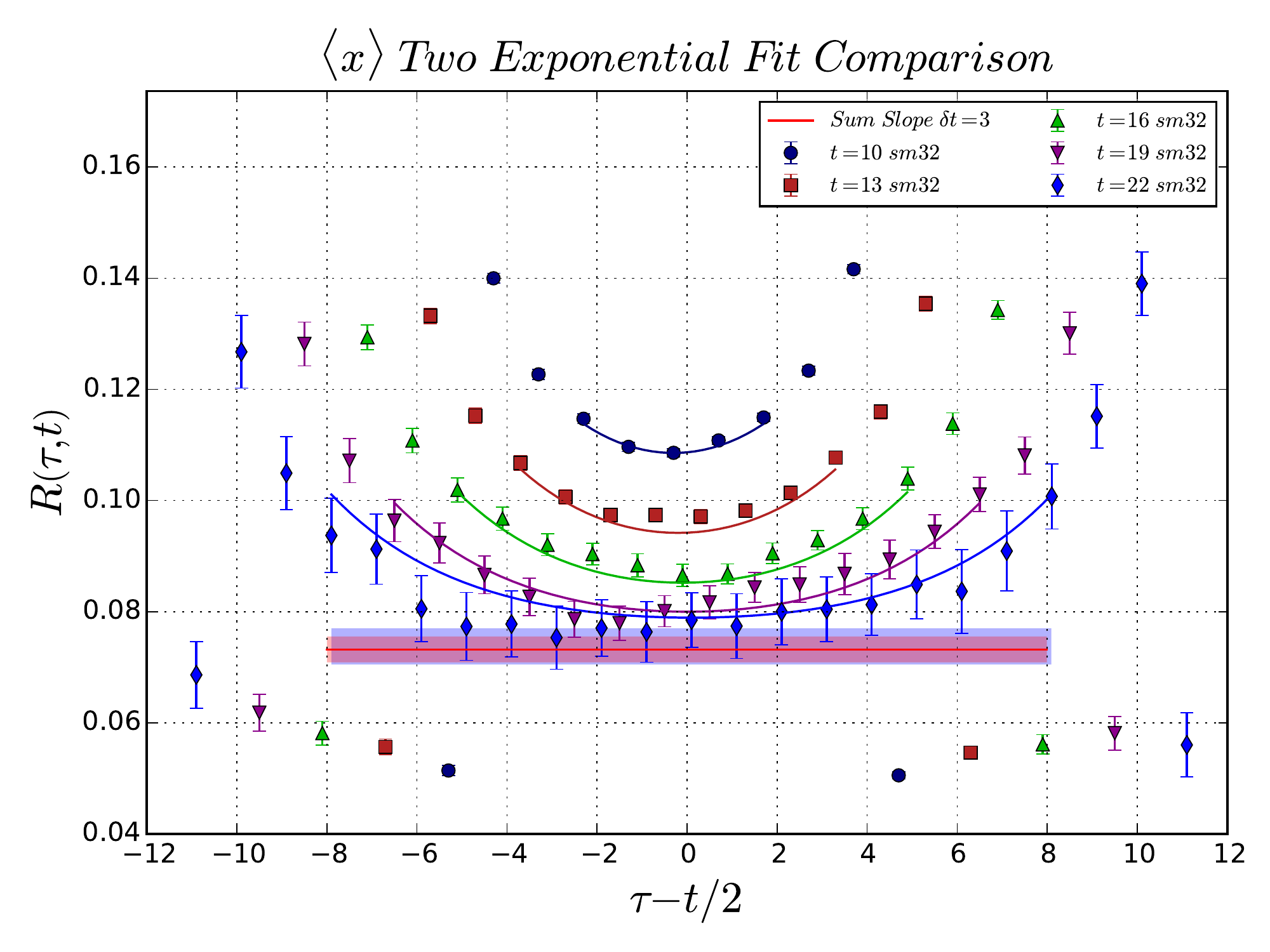}
\caption{\label{giDi2expTsinkGraph}Graph for \(\momfrac\) extracted at each current time \(\tau\) for multiple sink times plotted as a comparison to the summation method using a \(\delta t = 3\) (purple line in Figure~\ref{giDiSumGraph}). Also overlaid with a single two-exponential fit over both current and sink time (\(t\) and \(\tau\)). The lines correspond to the two-exponential fit function constructed and the blue shaded area corresponds to the \(\momfrac\) parameter extracted from the two-exponential fit. All results are calculated with 32 multiples of smearing at the source and sink.}
\end{figure}

The two-exponential fit to all 5 source-sink time separations for \(\momfrac\) in Figure~\ref{giDi2expTsinkGraph} has been more successful relative to the previous two quantities. We see the fit function being approximated appropriately for all current time and source-sink data sets. But as discussed in the summation method, we must be sure that the two-exponential approximation is satisfied, especially as the excited state contamination is so large for \(\momfrac\).

\begin{figure}
\includegraphics[trim={4mm 5mm 4mm 3mm},clip,width=.48\textwidth]{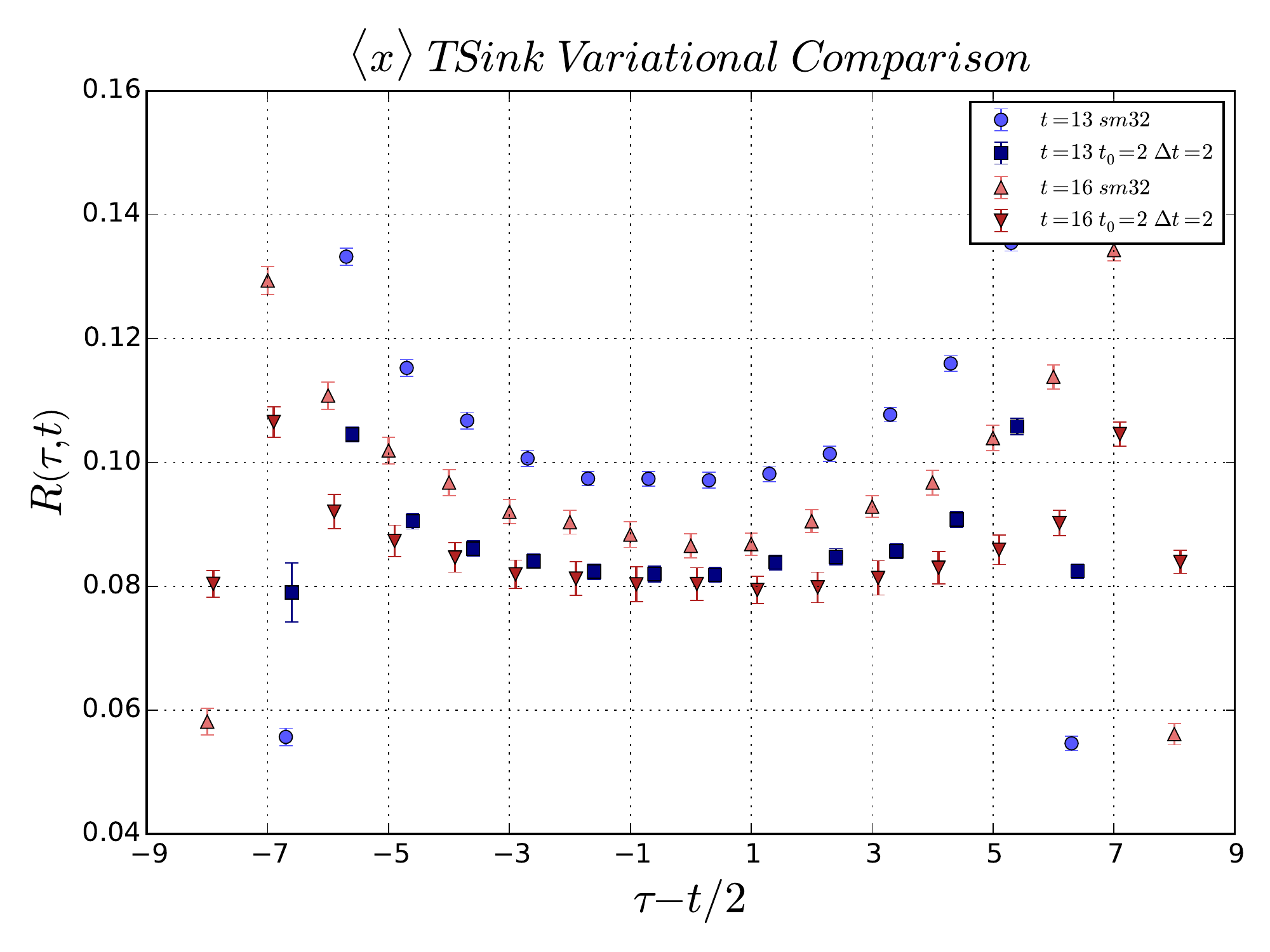}
\caption{\label{giDiTSinkVarGraph}Graph for \(\momfrac\) extracted at each current time \(\tau\) comparing different variational results for different source-sink separations (square and up-side-down triangle points). This is overlayed with lighter coloured results which is the corresponding source-sink separation for the 32 sweeps of smearing result (circle and triangle points).}
\end{figure}

In the case of \(\momfrac\), as displayed in Figure~\ref{giDiTSinkVarGraph}, we see no statistically significant
difference between the variational method for the two source-sink
separations which implies the variational method has dramatically reduced the amount of excited state contamination.
The same cannot be said about the single-smearing analysis.

\begin{figure}
\includegraphics[trim={4mm 5mm 4mm 3mm},clip,width=.48\textwidth]{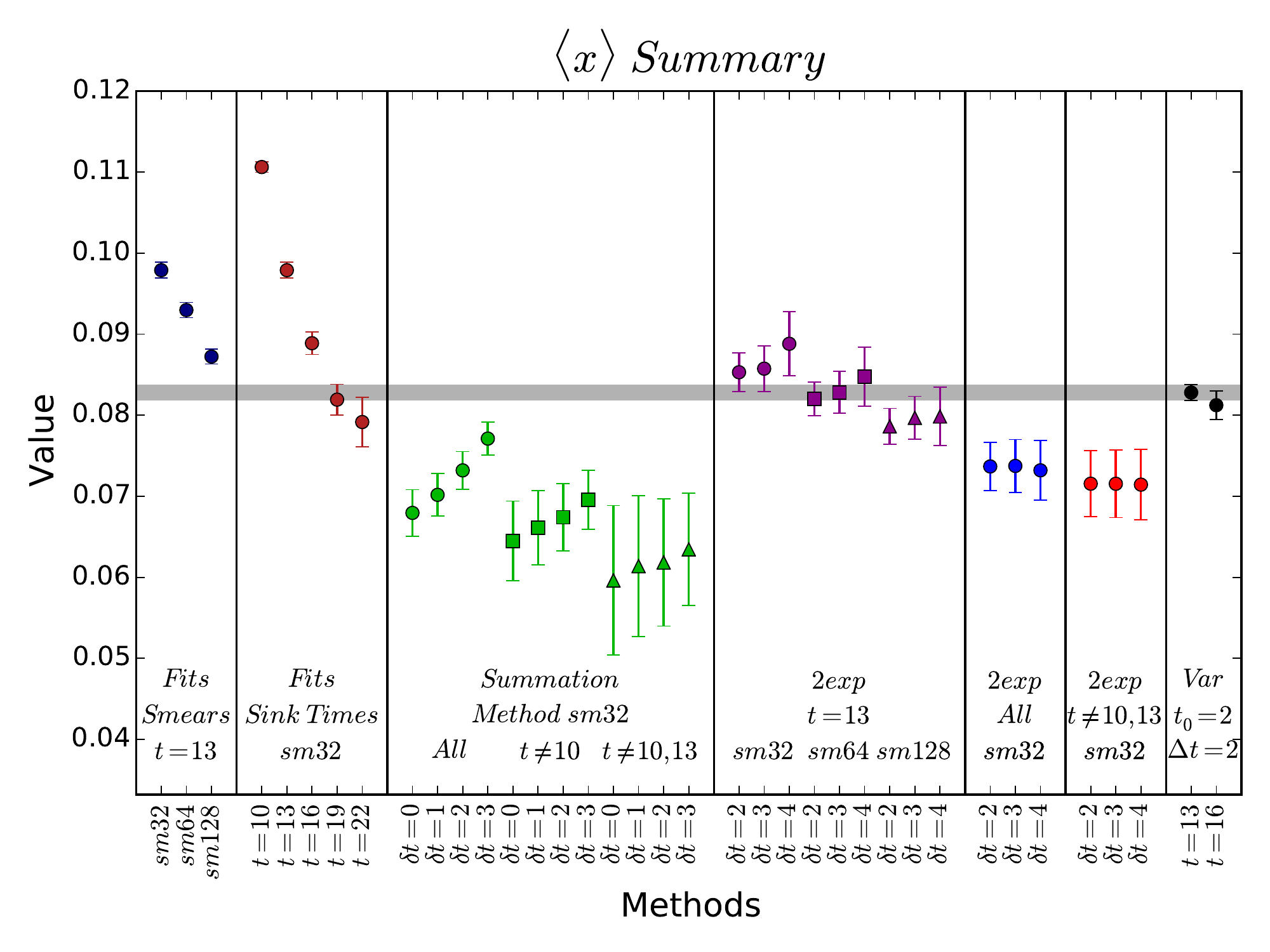}
\caption{\label{giDiSummary}Summary of all the extracted values for \(\momfrac\) over the different methods. ``Fits'' and ``Var'' use a constant fit range to extract a value. \(\delta t\) is defined in Eq.(\ref{eq:Sum}) for the summation method and corresponds to how many points are excluded from the source and sink current times in the two-exponential fit range. ``All'' refers to using all the source-sink separated results in the analysis and \(t\ne n\) refers to excluding n in the calculation. The shaded band helps to compare the variational method \(t=13\) result.}
\end{figure}

In summary for \(\momfrac\) (Figure~\ref{giDiSummary}) we see that the amount of excited state contamination removed by the variational method is at the point where the statistical precision has become a larger factor. This puts into question the validity of the summation method and the two-exponential fit results as they show a large disagreement to the variational method. This could be due to insufficient source-sink separated values skewing the results as is indicated by the summation method having a \(\delta t\) dependence when it should not. The larger uncertainties due to using very large source-sink separated results could also contribute to the disagreement.

\begin{table} 
\begin{ruledtabular}
\begin{tabular}{l|l|l|l}
Methods& \(g_{A}\) & \(g_{S}\) & R for \( \momfrac\)\\ \hline
Fit t=13, sm32                  	& 1.0524(90)	& 0.829(21) 	& 0.09790(98)\\
Fit t=13, sm64        			& 1.0727(82)	& 0.871(19) 	& 0.09298(94)\\
Fit t=13, sm128    			& 1.1009(86)	& 0.922(20) 	& 0.08724(91)\\ \hline
Fit t=10, sm32     			& 1.0047(52)	& 0.733(12) 	& 0.08724(91)\\
Fit t=13, sm32     			& 1.0524(90)	& 0.829(21) 	& 0.09790(98)\\
Fit t=16, sm32     			& 1.079(15)	& 0.896(34) 	& 0.0889(14)\\
Fit t=19, sm32     			& 1.079(26)	& 0.956(53) 	& 0.0819(19)\\
Fit t=22, sm32     			& 1.098(45)  	& 0.975(90) 	& 0.0792(31)\\ \hline
Sum All \(\delta\)t=0     		& 1.145(27) 	& 1.034(78) 	& 0.0680(28)\\
Sum All \(\delta\)t=1     		& 1.136(25) 	& 1.016(72) 	& 0.0702(26)\\
Sum All \(\delta\)t=2   		& 1.127(23) 	& 0.994(65) 	& 0.0732(23)\\
Sum All \(\delta\)t=3   		& 1.115(21) 	& 0.965(57) 	& 0.0771(20)\\
Sum t\(\ne\)10 \(\delta\)t=0   	        & 1.119(51) 	& 1.12(13) 	& 0.0645(49)\\
Sum t\(\ne\)10 \(\delta\)t=1   	        & 1.117(48) 	& 1.10(12) 	& 0.0661(46)\\
Sum t\(\ne\)10 \(\delta\)t=2   	        & 1.113(45) 	& 1.07(11) 	& 0.0674(41)\\
Sum t\(\ne\)10 \(\delta\)t=3   	        & 1.109(42) 	& 1.05(10) 	& 0.0696(36)\\
Sum t\(\ne\)10,13 \(\delta\)t=0   	& 1.10(10) 	& 1.28(22) 	& 0.0596(92)\\
Sum t\(\ne\)10,13 \(\delta\)t=1   	& 1.105(99) 	& 1.22(20) 	& 0.0614(87)\\
Sum t\(\ne\)10,13 \(\delta\)t=2   	& 1.104(94) 	& 1.17(19) 	& 0.0614(87)\\
+Sum t\(\ne\)10,13 \(\delta\)t=3   	& 1.102(87) 	& 1.13(17) 	& 0.0635(69)\\ \hline
2exp t=13, sm32 \(\delta\)t=2  	        & 1.121(24) 	& 0.961(38) 	& 0.0853(23)\\
2exp t=13, sm32 \(\delta\)t=3  	        & 1.125(25) 	& 0.969(39) 	& 0.0857(28)\\
2exp t=13, sm32 \(\delta\)t=4     	& 1.125(26) 	& 0.985(41) 	& 0.0888(39)\\
2exp t=13, sm64 \(\delta\)t=2     	& 1.115(22) 	& 0.979(36) 	& 0.0820(21)\\
2exp t=13, sm64 \(\delta\)t=3     	& 1.117(22) 	& 0.981(37) 	& 0.0828(26)\\
2exp t=13, sm64 \(\delta\)t=4     	& 1.116(22) 	& 0.990(38) 	& 0.0848(37)\\
2exp t=13, sm128 \(\delta\)t=2     	& 1.126(26) 	& 1.013(40) 	& 0.0786(22)\\
2exp t=13, sm128 \(\delta\)t=3     	& 1.125(26) 	& 1.011(41) 	& 0.0797(27)\\
2exp t=13, sm128 \(\delta\)t=4          & 1.123(26) 	& 1.015(41) 	& 0.0799(36)\\ \hline
2exp All sm32 \(\delta\)t=2             & 1.087(35) 	& 0.974(59) 	& 0.0737(30)\\
2exp All sm32 \(\delta\)t=3             & 1.093(37) 	& 0.981(66) 	& 0.0737(33)\\
2exp All sm32 \(\delta\)t=4             & 1.096(42) 	& 0.996(82) 	& 0.0732(37)\\ \hline
2exp t\(\ne\)10,13 sm32 \(\delta\)t=2  	& 1.090(47) 	& 1.03(11) 	& 0.0716(41)\\
2exp t\(\ne\)10,13 sm32 \(\delta\)t=3	& 1.094(48) 	& 1.02(11) 	& 0.0715(42)\\
2exp t\(\ne\)10,13 sm32 \(\delta\)t=4 	& 1.095(49) 	& 1.02(12) 	& 0.0714(43)\\ \hline
Var t=13, \(t_{0}\)=2 \(\Delta\)t=2 	& 1.1203(95)	& 0.963(23) 	& 0.08281(97)\\
Var t=16, \(t_{0}\)=2 \(\Delta\)t=2 	& 1.118(16)	& 0.942(47) 	& 0.0812(18)\\
\end{tabular}
\end{ruledtabular}
\caption{\label{tab:Results}Summary of results as displayed in Figures \ref{gASummary}, \ref{ISummary}, \ref{giDiSummary}. To obtain \(\momfrac\), one must divide out the mass of the proton and then apply the renormalisation constant to R.}
\end{table}

\section{\label{Ana}Summary and discussion}

A table of our results for \(g_{A}\), \(g_{S}\) and R for \(\momfrac\) presented in the previous section is given in Table \ref{tab:Results}. Here we summarise our findings.

\subsection{\label{Ana:Sum}Summation Results}
In Figures \ref{gA2expTsinkGraph}, \ref{I2expTsinkGraph}, \ref{giDi2expTsinkGraph}, we observe that the summation method looks as if it is improving the result. However, when looking at \(g_{S}\) and R for \(\momfrac\) extracted values in their respective summary plots (Figures \ref{ISummary}, \ref{giDiSummary}) we can see a dependence in the \(\delta t\) value when, if our two-exponential ansatz were satisfied, it should have no or minimal effect.

This is seen more clearly when considering summation fits excluding smaller source sink separations (\(g_{A}\) in Figure~\ref{gASumGraph}, \(g_{S}\) in Figure~\ref{ISumGraph} and \(\momfrac\) in Figure~\ref{giDiSumGraph}). When we exclude the smaller source-sink separated results, we can see that the two-exponential ansatz is breaking down for \(g_{S}\) and \(\momfrac\) as the data points do not lie within the errors projected to earlier source-sink separated time values.

\subsection{\label{Ana:TSF}Two-Exponential Fit Results}
The ``Two Exponential Variational Comparison'' plots seem to show minimal improvement for \(g_{A}\) (Figure~\ref{gA2expVarGraph}), some improvements for \(g_{S}\) (Figure~\ref{I2expVarGraph}) and the most improvement for \(\momfrac\) (Figure~\ref{giDi2expVarGraph}). Poor determination would be attributed to not being able to distinguish excited state contamination from our error within a fit range in which a two-exponential ansatz is justified. These results give a good demonstration of using fitting functions to remove transitional matrix elements. In all cases, the smaller smeared results (with larger excited state contamination) extract a value closer to the larger smeared results. From the summary plots (Figures \ref{gASummary}, \ref{ISummary}, \ref{giDiSummary}), we see minimal effect on the \(\delta t\) fit parameter for the two-exponential fit method.

Extending to the full source-sink separated set of results in ``Two Exponential Fit Comparison'' for 32 sweeps of smearing (Figures \ref{gA2expTsinkGraph}, \ref{I2expTsinkGraph}, \ref{giDi2expTsinkGraph}), we see that the fit is weighted predominately by the smallest source-sink separations. Furthermore, we see how poorly the larger source-sink separated results are in terms of symmetry about the middle current insertion time, as well as deformations to the expected curved fit lines. Although using the two-exponential fit method controls the excited states better than using a single source-sink separation, we found there was no improvement to a constant fit over the largest source-sink separation for \(g_{A}\) and \(g_{S}\) and a questionable improvement for \(\momfrac\).

\subsection{\label{Ana:VarTSink}Variational Results}
Beginning with the effective mass plots in Figure~\ref{MassGraph} where the effective masses for the three different smearing results were compared to the variational method, the variational method allows us to extract the mass from the two-point correlator beginning from an earlier time slice compared to the individually smeared results. The improvement is due to the excited states being suppressed when constructing the optimal correlator in Eq.\ref{eq:CM2ptConst}.

In Figures \ref{gAGraph}, \ref{IGraph}, \ref{giDiGraph} we compare the ratio functions (Eq.\ref{eq:Rfun}) for the three different smearing results to the variational method in which the functions are varied over the current insertion time \(\tau\) for a fixed source-sink separation \(t=13\). The figures show how applying the variational method improves the suppression of excited state contamination. The ability to fit a plateau over a much larger current insertion time \(\tau\) shows how the transition matrix elements are being sufficiently suppressed compared to the individually smeared results. The shift in each of the ratio values for each particular \(\tau\) shows how the variational method is suppressing all types of excited state contamination (``transition'' and ``excited to excited state'' matrix elements).

The final collection of graphs ``TSink Variational Comparison'' (Figures \ref{gATSinkVarGraph}, \ref{ITSinkVarGraph}, \ref{giDiTSinkVarGraph}) compares the variational method to the 32 sweeps of smearing results over the current insertion times \(\tau\) and the source-sink separation of 13 and 16. All 3 quantities calculated with the variational method show no statistically significant difference between the two source-sink separations. This shows us that choosing a source-sink separation of 13 for the variational method gives us a result where the residual excited state contamination is smaller than the errors. Compared to the tinted points (circle and triangle points), a much larger source-sink separation in the 32 sweeps of smearing case is needed to remove the remaining excited state contamination.





\subsection{\label{Dis:Findings}Findings}
We can see that in all values analysed, the variational method improved our result with only sacrificing minimal uncertainty. Varying the variational parameters showed to be irrelevant as all variations were consistent with each other.

In contrast, the summation and two-exponential fit methods either fell short of removing the excited state contamination or required the inclusion of source-sink time separations that induced large uncertainties in the results. Also, careful consideration must be taken to the two-exponential ansatz in both methods, as using insufficient source-sink separations might not satisfy the ansatz for any of the current insertion times. The two-exponential fit will improve as you improve the statistics of the calculation, as you will be able to distinguish the ground and excited state better from the uncertainties on the values. A possible improvement might be to weight the larger source-sink separated results with more statistics over the shorter source-sink separated results.

\subsection{\label{Dis:CBAna}Cost/Benefit Analysis}
\begin{table}[!t] 
\begin{ruledtabular}
\begin{tabular}{l|l|l|l}
Create& Standard & 2exp \& SM (over \(n_{t}\)) & CM (over \(n_{basis}\))\\ \hline
\(C_{2}\) & \(1\) & \(1\) & \(n_{basis}\)\\
\(C_{3}\) & \(4\)& \(4n_{t}\)& \(4n_{basis}\)\\
Total & \(5\) & \(1+4n_{t}\) & \(5n_{basis}\)\\
This Paper & \(5\) & \(21\) & \(15\)
\end{tabular}
\end{ruledtabular}
\caption{\label{tab:comp}Comparative computational times for each method for two-point and three-point correlators as well as the total. \(n_{t}\) is the number or source-sink time separations used and \(n_{basis}\) is the number of basis interpolating fields used.}
\end{table}

Assuming we have an equal number of gauge fields for our particular \(\kappa\) value (or pion mass), we can model the efficiency as to how many inversions we undertake per gauge field. One inversion is required for calculating the two-point correlator, then a second inversion is required for each specific three-point correlator we want to calculate. The fixed sink method requires that we choose a sink time, sink momentum, spin projector and which quark the current acts on for a fixed hadron before the three-point correlator is calculated.

The variational method requires \(n_{basis}\) inversions to create the two-point correlators, where  \(n_{basis}\) is the number of basis interpolating fields used (e.g. 3 smearings for this work). Then a further \(n_{basis}\) is required to create a particular fixed sink resulting correlator as shown in Section \ref{VSm}.

The two-exponential fit and summation methods are identical to the standard way, but creating \(n_{t}\) multiples of the three-point correlator, where \(n_{t}\) is the number or source-sink time separations.

For this analysis, simulations were performed with zero sink momentum and two different spin projectors for both up and down quark contributions to the proton. This results in 4 times the number of inversions for each three-point correlator required. The inversion numbers are outlined in Table \ref{tab:comp}.

\section{\label{Con}Conclusion}
In lattice simulations of three-point correlation functions it is most common to make use of a sequential inversion ``through the sink''. This allows the efficient study of many operators and choices of momentum transfer for essentially fixed computational cost. To gain control of statistical uncertainties, it is preferable to keep the source-sink separation time short. Unfortunately, aggressive choices of source-sink separations leads to significant contamination from excited states. One can extend the source-sink separation, yet for fixed computational cost, the results presented here suggest that by the time the excited-state contamination is under control the statistical signal is almost lost. This motivates the study of competing techniques which have been proposed to mitigate the excited-state contamination problem.

Theoretically, the summation method offers a parametric suppression of excited-state contamination. Never the less, in similar fashion to the plateau method, we find this technique to be plagued by the difficulty of identifying the shortest source-sink separation which can reliably be used in a given fit. The high statistical precision obtained at short source-sink separated times can potentially lead to a significant distortion of the fit and result in erroneous extraction of matrix elements.

The two-exponential fit allows the influence of excited-state contamination to be accounted for numerically. The analysis presented here suggests that this technique offers an improved determination of the desired matrix elements. The method appears rather robust with respect of modified fit ranges, which might indicate that the two exponentials are sufficient to model the two states of the correlators. The uncertainty estimate appears reliable in general, yet caution should be taken if the extracted value lies outside the fit at the largest source-sink time separation.

In contrast to the two previous techniques, which require investigation of an extended range of source-sink separated correlators, the variational approach is designed to reduce the excited state contamination at early times where the statistical signal is still strong. We find that we were reliably able to apply a plateau fit to the variational method calculation due to obtaining a larger number of current insertion time results that had plateaued to a common value. This indicates that all transition matrix elements were sufficiently suppressed with respect to the uncertainties. Although we knew that all excited state contamination effects should be suppressed from examining the effective mass plots (Eq.\ref{MassGraph}), having a larger source-sink separated result for the variational method confirmed our initial choice of source-sink time separation.

We anticipate that the results presented here will be naturally applicable to a more general set of observables. In particular, at finite momentum transfer the variational approach can be easily adapted to allow for momentum-dependent operator projection at the source. Although a priori knowledge of a semi-optimal smearing for zero momentum operator projection at the source and sink may be sufficient for these types of calculations, moving to momentum-dependant operator projection at the source may have different optimal smearings for each source momentum calculated. Results will be presented in a future publication.

While the results presented here are just for a single quark mass, the issue of excited state contamination is anticipated to become even more prevalent at light quark masses and large volumes. Given that statistical fluctuations are also greater at light quark masses, there will be increasing demand for techniques which are robust at short source-sink separations, such as the variational method described here.



\section{\label{ack}Acknowledgements}
The generation of the numerical configurations was performed using the BQCD lattice QCD program, \cite{Nakamura:2010qh}, on the IBM BlueGeneQ using DIRAC 2 resources (EPCC, Edinburgh, UK), the BlueGene P and Q at NIC (J\"{u}lich, Germany) and the Cray XC30 at HLRN (The North-German Supercomputing Alliance). Some of the simulations were undertaken on the NCI National Facility in Canberra, Australia, which is supported by the Australian Commonwealth Government. We also acknowledge the Phoenix cluster at the University of Adelaide. The BlueGene codes were optimised using Bagel \cite{Boyle:2009vp}. This investigation has been supported in part by the Australian Research Council under grants FT120100821, FT100100005, DP150103164, DP140103067 and CE110001004.

\bibliography{JackJournal}

\end{document}